\documentclass[prb,twocolumn,superscriptaddress,
showpacs,amsmath,amssymb]{revtex4-1}

\usepackage{esint}
\usepackage{graphicx}
\usepackage{latexsym}
\usepackage{amsmath}
\usepackage{amssymb}
\usepackage{amsfonts}
\usepackage{color}
\usepackage{bm}
\usepackage{verbatim}

\definecolor{MS-color}{RGB}{128,0,128}

\usepackage{framed}
\definecolor{shadecolor}{RGB}{222,222,221}
\begin{document}

    \title{Finite-frequency spin susceptibility and spin pumping in superconductors with spin-orbit relaxation}

\date{\today}

 \author{M.A.~Silaev}
 \affiliation{Department of
Physics and Nanoscience Center, University of Jyv\"askyl\"a, P.O.
Box 35 (YFL), FI-40014 University of Jyv\"askyl\"a, Finland}
\affiliation{Moscow Institute of Physics and Technology, Dolgoprudny, 141700 Russia}
\affiliation{Institute for Physics of Microstructures, Russian Academy of Sciences, 603950 Nizhny Novgorod, GSP-105, Russia}
%
 \begin{abstract}
 Static spin susceptibility of superconductors with spin-orbit
 relaxation has been calculated in the seminal work of A.A. Abrikosov
 and L.P. Gor'kov [Sov. Phys. JETP, {\bf 15}, 752 (1962)]. Surprisingly the generalization of this result to   finite frequencies has not been done despite being quite important for the 
modern topic of superconducting spintronics.  
 The present paper fills this gap by deriving the analytical expression for spin susceptibility. The time-dependent spin response is shown to be captured by the quasiclassical Eilenberger equation with  collision integrals corresponding to the ordinary and spin-orbit scattering. Using the developed formalism we study the linear spin pumping effect between the ferromagnet and the adjacent superconducting film.  The consequences for understanding recent experiments demonstrating the modification of Gilbert damping by the superconducting correlations are discussed.  
  \end{abstract}

\pacs{} \maketitle

\section{Introduction}
 
 Spin transport and spin dynamics in superconductors have attracted significant attention recently\cite{Linder2015,RevModPhys.90.041001, han2019spin,quay2018out,ohnishi2020spin,beckmann2016spin,Eschrig2015a}. 
Quite interesting experimental results have been obtained 
for the spin pumping effects which in general play the central role in  spintronics \cite{brataas2002spin,tserkovnyak2002enhanced,RevModPhys.77.1375}. 
 Ferromagnet/ 
 superconductor multilayers  were found recently to demonstrate  changes
 of the ferromagnetic resonance (FMR) frequency and linewidth \cite{bell2008spin,jeon2019abrikosov,PhysRevB.99.024507,PhysRevApplied.11.014061,Jeon2018,yao2018probe,li2018possible,zhao2020exploring,golovchanskiy2020magnetization} due to the superconducting correlations. 
  %
 Despite significant efforts theoretical understanding of these effects is not complete yet. 
 For example, puzzling experimental result has been obtained for the ferromagnetic insulator/superconductor multilayers where the pronounced peaks in the temperature dependence of Gilbert damping have been observed \cite{yao2018probe}. 

The enhancement of Gilbert damping due to the metal spin sink 
 can be  calculated using the  linear response approximation \cite{ohnuma2014enhanced}  
 which involves the 
 momentum and frequency-dependent spin susceptibility  $\chi_h(k,\Omega)$  
 of the  metal spin sink. 
 %
  Hence, to understand the modification of Gilbert damping due to the spin pumping in superconducting films it is necessary to the  calculate
  the corresponding function $\chi_h(k,\Omega)$  in the presence of spin relaxation mechanisms like the spin-orbit scattering.
 Quite surprisingly, this calculation has not been ever performed correctly. Recent papers which have addressed this topic in connection with spin pumping \cite{inoue2017spin, taira2018spin} report finite 
zero-temperature dissipation at low frequencies: $ {\rm Im} \chi_h(q,\Omega)/\Omega \neq 0$ at $\Omega \to 0$. 
  This result contradicts physical intuition because there can be no dissipation at $\Omega<2\Delta$ and in the absence of thermal quasiparticles which are frozen out in superconductors at $T\ll \Delta$, where $\Delta$ is the superconducting energy gap. As we show below this inconsistency  comes from neglecting the important contributions while performing analytical continuation procedure. 
   
    The first purpose of the present paper is to report the analytical expression for the finite-frequency spin susceptibility of superconductors with spin-orbit relaxation mechanism. This result is a generalization of the classical work of Abrikosov and Gor'kov \cite{abrikosov1962spin} who have considered the static spin susceptibility to explain the finite Knight shift in superconductors at $T\ll \Delta$. We analyse different characteristic regimes including large and strong spin relaxation as well as the behaviour for various values of the Dynes parameter\cite{Dynes1984}.
    
   The second purpose is to study the spin pumping in superconductor/ferromagnet systems in the framework of the interfacial exchange model \cite{ohnuma2014enhanced}. The expressions for Gilbert damping are derived for the finite thickness of the spin sink layer. Also we consider the system with  an additional  perfect spin absorber which can be realized experimentally by adding the layer of material with very strong spin relaxation. The derived general expressions can be parametrized in terms of the dimensionless parameter characterizing the strength of the interfacial coupling between the ferromagnet and adjacent superconductor. Systems with elevated values of this parameter are predicted to feature pronounced shift of the ferromagnetic resonance line induced by superconducting correlations.  
        
   

 \begin{figure}[htb!]
 \centerline{$
 \begin{array}{c} 
   \includegraphics[width=0.27\linewidth]{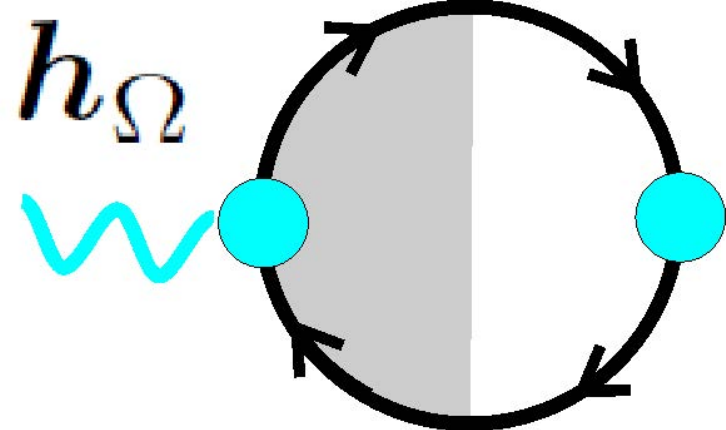} 
   \put (-100,30) {\Large {\color{black} (a) }}
  \\
   \includegraphics[width=1.0\linewidth]{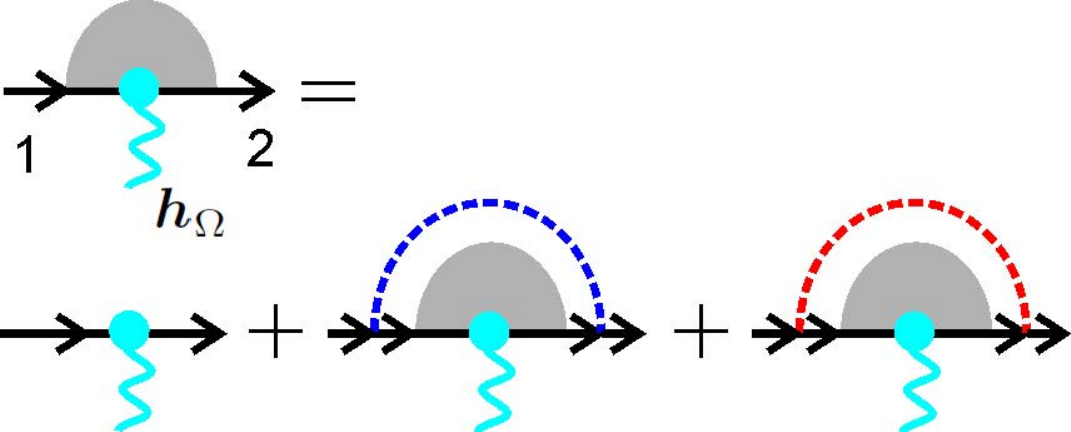}
   \put (-250,90) {\Large {\color{black} (b) }}
   \end{array}$}
 \caption{\label{Fig:Bubble} (Color online) 
 (a) Bubble diagram for the linear response of spin polarization 
 generated by the time-dependent Zeeman field $\bm h_\Omega  e^{i\Omega t + i\bm q\bm r}$ shown by the wavy line. Circles show spin vertices $\hat {\bm \sigma}$. The shaded region shows  impurity ladder.
 (b) Diagrammatic equation for the impurity ladder.  The blue and red dashed lines correspond to the ordinary and spin-orbit scattering potentials averaged over the random impurity configuration. 
  }
 \end{figure}   
   
   
   {
 \section{General formalism }
    \subsection{Diagrammatic formalism}
 We describe the interaction of electrons with 
 Zeeman  field $\bm h= \bm h (\bm r, t)$
 using the following Hamiltonian
 \begin{align} \label{Eq:Hp}  
 & \hat V_P = \hat{ \bm\sigma}  \bm h  
 \end{align}      
 where 
 $\hat{ \bm\sigma}  = (\hat\sigma_x, \hat\sigma_y, \hat\sigma_z )$ is the vector of spin Pauli matrices. 
  Besides that we assume the presence of disorder described by the 
 Gaussian impurity potential. It has both the usual  
  $V_{imp}$  and the spin-orbit $V_{so}$ scattering amplitudes\cite{}
  \begin{align}
 & \hat V (\bm p, \bm p^\prime) = 
  \\ \nonumber
 & u_0  
 \sum_{\bm r_{o}} e^{ i \bm r_{o}(\bm p - \bm p^\prime) } 
 +
  \frac{u_{so}}{p_F^{2}} \hat {\bm\sigma} 
 \cdot (\bm p \times \bm p^\prime)  
 \sum_{\bm r_{so}} e^{ i \bm r_{so}(\bm p - \bm p^\prime) } ,
  \end{align}
 where $\bm r_o$ and $\bm r_{so}$ denote the random impurity coordinates corresponding to the ordinary and spin-orbit scattering respectively. We assume this coordinates to be independent and thus  neglect the magnetoelectric effects arising from the combined ordinary and spin-orbit scattering \cite{bergeret2016manifestation}. 
      
   The spin polarization as a function of the imaginary time 
      $ t\in [0, \beta]$ where $\beta=1/T$ is given by  
   \begin{align} \label{Eq:SpinGFgeneral}
   & \bm S (t,\bm r) =  
   \frac{1}{4} {\rm Tr} [ \hat{ \bm\sigma}   
   \hat G ](\bm r,\bm r, t_{1,2}=t)  
    \end{align}
 where $\hat G (\bm r_1,\bm r_2, t_{1,2})$ is the imaginary time 
 Green's function (GF).
  The stationary propagators depend only on the relative time and coordinate. In the frequency and momentum representation they are 
 given by 
 \citep{abrikosov2012methods,abrikosov1962spin} 
 \begin{align} \label{Eq:GFimp}
 & \hat G_0 (\omega,\bm p)= \frac{ \tilde\Delta \hat\tau_2 -
 i\tilde{\omega} 
 \hat\tau_0 + \xi_p \hat\tau_3}
 {\tilde\Delta^2 + \tilde{\omega}^2 + \xi_p^2}
 \\
 & \tilde{\omega}= \omega \frac{\tilde{s} (\omega) }{s(\omega)};
 \;\; \tilde{\Delta}= \Delta \frac{\tilde{s} (\omega) }{s(\omega)} , 
 \end{align}
 where $\xi_p = p^2/2m - \mu$ is the deviation of the kinetic  
 energy from the chemical potential $\mu$ and 
 $\hat \tau_{1,2,3}$ are the Pauli matrices in Nambu space. 
 We denote $s= \sqrt{\omega^2 + \Delta^2}$ and $\tilde{s} = s + 1/2\tau_{imp}$ where the scattering time is given by the superposition
  $\tau_{imp}^{-1} = \tau_{o}^{-1} + \tau_{so}^{-1} $. We denote the usual 
  $\tau_o^{-1} = 2\pi n \nu  u_0  $   
  and spin-orbit
  $\tau_{so}^{-1} = 2\pi n \nu  u_{so}/3 $ scattering rates.
 The 
 propagator (\ref{Eq:GFimp}) is averaged over the randomly 
 disordered point scatterers configurations. 
        
  We are interested in the spin polarization induced by the 
 external Zeeman field $\bm h (t, \bm r) = \bm h_\Omega e^{i\Omega t + i\bm q\bm r}$.  The induced spin polarization as given by the  diagram shown in Fig.\ref{Fig:Bubble}a can be written as follows
 \begin{align} \label{Eq:Susceptibility0}
 \bm S_\Omega = \chi_h (\Omega, \bm q) \bm h_\Omega
 \end{align}  
  The linear spin susceptibility is defined by 
 substituting into the Eq.\ref{Eq:SpinGFgeneral}   
  $\hat G_h$ which is the  first-order correction to the GF induced by the Zeeman field. 
  The diagrammatic equation for this correction which includes the summation of impurity ladder corrections is shown Fig.\ref{Fig:Bubble}b.
 The shaded region denotes impurity ladder corresponding to the 
 ordinary and  spin-orbit impurity scattering averaged over 
 the random disorder configuration.
  The red and blue dashed lines correspond 
  to the spin-orbit and ordinary impurity scattering potentials averaged over the randomly distributed impurities.  
    }
      Analytical expression for the diagrammatic equation in Fig.\ref{Fig:Bubble}b reads as follows 
 \begin{align} \label{Eq:IntegralG_h}
 & \hat G_{h}(12) = 
 - \hat G_{0}(1)
 \hat {\bm\sigma}  \bm h_\Omega
 \hat G_{0}(2) 
 +
 \\ \nonumber
 & \hat G_{0}(1) 
 \frac{\hat {\bm\sigma}  \langle\hat g_h\rangle \hat {\bm\sigma}  \hat\tau_3}
 {6i\tau_{so}}
 \hat G_{0}(2)
 +
\hat G_{0}(1) 
 \frac{ \langle\hat g_h\rangle  \hat\tau_3}
 {2i\tau_{o}}
 \hat G_{0}(2)
 \end{align}     
 %
 where we have introduced the notation 
 \begin{align} 
 \hat g_{h}=\frac{i}{\pi}
  \fint d\xi_p \hat\tau_3  \hat G_{h} .
 \end{align}
 We use the condensed notation $\hat G_0(2) =  \hat G_0(\omega, \bm p)$ and $\hat G_0(1) =  \hat G_0(\omega - \Omega, \bm p + \bm q)$. 
 The correction depends on the two frequencies and momenta $\hat G_{h}(12) = \hat G_{h}(\omega_1,\bm p, \omega_2, \bm p + \bm q)$.
 The angular brackets denote average over the momentum directions on the 
 Fermi sphere so that in total 
 $\langle\hat g_h\rangle = (i/\pi \nu) \int d^3p \hat G_h$, where
 $\nu$ is the density of states at the Fermi level.  
 Diagrammatically the equation for impurity ladder (\ref{Eq:IntegralG_h}) is shown in Fig.\ref{Fig:Bubble}b. 
 The second and third terms in Eq.\ref{Eq:IntegralG_h} corresponding to the spin-orbit and ordinary scattering are shown 
 by blue and red dashed lines, respectively. 
 As we see below the momentum-integrated correction $\hat G_h$ coincides with the solution of quasiclassical 
 Eilenberger equation \cite{Eilenberger1968} with collision integrals corresponding to the ordinary and spin -orbit scattering \cite{Bergeret2005a}. 
 
 
  { \subsection{Quasiclassical formalism}
  
  Under quite general conditions the non-equilibrium state of a metal involves perturbations of spectrum and distribution function in the vicinity of the Fermi level. 
  For that the external fields should have frequencies much smaller than the Fermi energy and spatial scales much larger than the Fermi wave length. Both these requirements are satisfied for the spin pumping systems. Hence we can use the theory formulated in terms of the quasiclassical propagator \cite{Eilenberger1968}
 \begin{align} \label{Eq:QuasiclassicalG}
 \hat g (\bm r, \bm n_p, t,t^\prime)=  \frac{i}{\pi} \fint 
 d\xi_p  \hat\tau_3\hat G.
 \end{align}

 The calculation can be performed either using the imaginary time 
 formalism of the real-time formalism. 
 {
 In the imaginary time domain the quasiclassical propagator 
 is determined by the Eilenberger equation with collision integrals
 describing the impurity scattering\cite{Eilenberger1968}
 \begin{align} \label{Eq:Eilenberger} 
 & (\bm v_F \bm \nabla) \hat g  - i\{ \hat
 \tau_3\partial_t , \hat g \}_t
 = 
  i[\hat\tau_3 \hat H, \hat g]_t + 
 [(\hat \Sigma_o + \hat \Sigma_{so})\circ, \hat g ]_t
  \\
 & \hat \Sigma_{so} = 
 (\hat {\bm\sigma}  \langle \hat g\rangle \hat {\bm\sigma}  ) /6\tau_{so} 
 \\    
 & \hat \Sigma_o =  \langle \hat g\rangle/2\tau_{o}  .
 \end{align}  
 Here  $\hat \Sigma_o$ and  $\hat \Sigma_{so}$ are the self-energies corresponding to the ordinary and spin-orbit scattering, respectively\cite{bergeret2005odd} and $\hat H =  \Delta \hat\tau_2 + \hat {\bm\sigma} \bm h $. 
 We denote the commutators 
 $[X,g]_t= X(t_1) g(t_1,t_2) - 
 g(t_1,t_2)X(t_2)$ 
 and the convolution 
 $\langle \hat g\rangle\circ \hat g  = 
 \int_0^\beta dt \langle \hat g\rangle(t_1,t) 
 \hat g (t,t_2) $. 
 The angle-averaging over the Fermi surface is given by 
 $\langle g\rangle$. 
 The spin polarization is given by 
 \begin{align} \label{Eq:CurrentQuasiclassic}
 & \bm S (t, \bm r) = -i  \frac{\pi  \nu}{4}  {\rm Tr} 
 [\hat\tau_3  \hat {\bm \sigma} \langle \hat g (t,t,\bm r) \rangle ]
            \end{align}
 The quasiclassical equations are supplemented by the normalization 
 condition $ \hat g\circ \hat g =1$.
 }

 }
  

   \subsection{Analytical continuation}
 \label{Sec:AnalyticalContinuation}
 In order to find the real-frequency response we need to implement 
 the analytic continuation of Eq.~\eqref{Eq:CurrentQuasiclassic}. 
 The first-order correction to the quasiclassical GF 
 can be written as  $\hat g_h (t_1,t_2) = T \sum_\omega e^{-i\omega_1 t_1 + i\omega_2 t_2} g(\omega_1,\omega_2)$ where $\omega_2= \omega$ and $\omega_1 = \omega - \Omega$ are the fermionic Matsubara frequencies shifted by the 
  Bosonic frequency  $\Omega$ of the external Zeeman field. 
   The analytic continuation of the sum  
 is determined according to the general rule \cite{kopnin2001theory}
  \begin{align} \label{Eq:AnalyticalContinuationGen}
  & T\sum_\omega g_h(\omega_1,\omega_2)
  \to  
  \\ \nonumber
  & \int \frac{d\varepsilon}{4\pi i} n_0(\varepsilon_1)
  \left[ g_h(-i \varepsilon^R_1,  -i \varepsilon^A_2) 
  - 
  g_h(-i \varepsilon^A_1,  -i \varepsilon^A_2)
  \right]
  +
  \\ \nonumber
  & \int \frac{d\varepsilon}{4\pi i}  
  n_0(\varepsilon_2)
  \left[ g_h(-i \varepsilon^R_1,  -i \varepsilon^R_2)
   - 
  g_h(-i \varepsilon^R_1,  -i \varepsilon^A_2)  \right]
     \end{align}
 where $n_0(\varepsilon) = \tanh (\varepsilon/2T ) $ is the equilibrium  
 distribution function. In the r.h.s. of (\ref{Eq:AnalyticalContinuationGen}) we substitute 
  $\varepsilon_1 = \varepsilon - \Omega$, 
  $\varepsilon_2 = \varepsilon$ 
  and  $\varepsilon^R = \varepsilon + i\Gamma$, $\varepsilon^A = 
  \varepsilon -i \Gamma$. Here the term with $\Gamma>0$ is added to   
  shift  the integration  contour into the corresponding half-plane.
  At the same time,  $\Gamma$ can be used as the Dynes 
  parameter \cite{dynes84}
  to describe the effect of different depairing mechanisms on spectral functions in the superconductor.
  We implement the analytical continuation in such a way that 
  $
  s(-i\varepsilon^{R,A}) 
  = - i \sqrt{ (\varepsilon^{R,A})^2- \Delta^2}
   $    assuming that the branch cuts run from $(\Delta,\infty)$
   and $(-\infty, -\Delta)$. 
 
 Equilibrium GF in the imaginary frequency domain is given by 
  $\hat g_0(\omega ) =  (\hat\tau_3 \omega + \hat\tau_1 \Delta)/s(\omega)$.
 The real-frequency continuation reads
  $\hat g^{R,A}_0(\varepsilon )  
 =  (\hat\tau_3 \varepsilon^{R,A} +i \hat\tau_1 \Delta)/\sqrt{ (\varepsilon^{R,A})^2- \Delta^2}$.

 Thus the linear response spin polarization is given by 
 \begin{align} \label{Eq:SpinAnalyticalContinuation}
 & \chi_h + 1= 
 \\ \nonumber
 & \int \frac{d\varepsilon}{4\pi i}
 \chi(-i \varepsilon^R_1,  -i \varepsilon^A_2) 
 \left[ n_0(\varepsilon_1) -  n_0(\varepsilon_2) \right] + 
  \\ \nonumber
 & \int \frac{d\varepsilon}{4\pi i}  
 \left[
  n_0(\varepsilon_2)
 \chi(-i \varepsilon^R_1,  -i \varepsilon^R_2)
 - 
 n_0(\varepsilon_1)
 \chi(-i \varepsilon^A_1,  -i \varepsilon^A_2)  \right]
 \end{align}  
 where we denote $\chi (\omega_1,\omega_2) = 
 (\delta/\delta \bm h){\rm Tr} [\bm \sigma \hat g_h (\omega_1,\omega_2)]$. 
 In the l.h.s. of Eq. \ref{Eq:SpinAnalyticalContinuation} we subtract the off-shell contribution to the spin polarization due 
to the band edge shift by the Zeeman field.  
 
 {It is interesting to note that in the superconducting state  both the first and the second terms in the r.h.s. of (\ref{Eq:SpinAnalyticalContinuation}) contribute to the dissipative part of spin susceptibility 
 With that we obtain physically correct behaviour in the low-temperature limit ${\rm Im} \chi_h(\Omega)/\Omega \to 0$ at $T\to 0$ and small frequency $\Omega \ll \Delta$. This is in contrast to previous calculations\cite{inoue2017spin, taira2018spin} which take into account only the first term in (\ref{Eq:SpinAnalyticalContinuation}) and obtained physically incorrect finite dissipation in the absence of quasiparticles at $T=0$. 
  }

  {
  \section{Spin susceptibility}
  \label{Sec:SpinSusceptibility}
    
 {
   
  \subsection{Diagram summation}
  \label{SubSec:1OrderCorr}

First, we demonstrate connection between response functions 
determined by the diagram Fig.\ref{Fig:Bubble}a and  by the solution of time-dependent Eilenberger equation (\ref{Eq:Eilenberger}).
Instead of using the usual approach of calculating the vertex function \cite{abrikosov1962spin} we use the alternative route and solve directly 
the equation for the first-order correction \ref{Eq:IntegralG_h}.

 We use the general approach suggested recently\cite{silaev2019nonlinear}
 for deriving equation for the momentum-integrated propagators $\hat g_h$  
 starting from the general equation for the exact GF (\ref{Eq:IntegralG_h}).
 The key idea of this derivation is based on the following trick. 
 Let us multiply 
 the function $\hat G_h (12)$ by $\hat G_0^{-1} (1)$ from the left
 and by $\hat G_0^{-1} (2)$ from the right, subtract the results 
 and integrate by $\xi_p$. 
 We use that  Eq.(\ref{Eq:GFimp}) yields the relations 
 $ \hat G_0^{-1} (j) = \tilde{\Delta}_j \hat\tau_2 
 + i \tilde{\omega}_j \hat\tau_0 + \xi_p(p_j) \hat\tau_3 $   
 and 
 $\tilde{\Delta}_j \hat\tau_2 + i \tilde{\omega}_j \hat\tau_0 = 
 i (s_j + 1/2\tau_{imp})\hat g_0(\omega_j)\hat\tau_3$.
 Then we eliminate off-shell contributions in the momentum integrals 
 to express the result through quasiclassical propagators 
 \begin{align} \label{Eq:lhs}
 & \int \frac{d\xi_p}{\pi} \left[ \hat G_0^{-1} (1) 
 \hat G_{h} - \hat\tau_3 \hat G_{h} 
 \hat G_0^{-1} (2) \hat\tau_3 \right]         
 = 
 \\ \nonumber
 & \tilde s_1\hat g_0(1) \hat g_h - 
 \tilde s_2\hat g_h \hat g_0(2)  - i(\bm v_F \bm q) \hat g_h
 \end{align}
  
        %
        %
Next let us derive the l.h.s. of the equation for $\hat g_h$. 
Using the diagram Fig.\ref{Fig:Bubble}b or the Eq.\ref{Eq:IntegralG_h} we get that 
           
   {
  \begin{align}
 &  \hat G_0^{-1} (1) 
 \hat G_{h} - \hat\tau_3 \hat G_{h}
 \hat G_0^{-1} (2)\hat\tau_3
 = 
 \\ \nonumber
 &
 \hat\tau_3\hat G_0(1)
 ( \hat{\bm  h}_{\Omega}  
 +
 i \langle \hat g_h\rangle \hat \tau_3/2\tau_{o} 
 + 
 i \bm\sigma \langle \hat g_h\rangle \bm\sigma 
 \hat \tau_3/6\tau_{so})\hat\tau_3
  \\ \nonumber
 & - ( \hat{\bm  h}_{\Omega}
  +
 i \langle \hat g_h\rangle \hat \tau_3/2\tau_{o} 
 +
  i \bm\sigma \langle \hat g_h\rangle \bm\sigma 
 \hat \tau_3/6\tau_{so}
   ) \hat G_0(2) 
 + 
  \end{align}
 where we denote $\hat{\bm  h}_{\Omega}   = \hat{ \bm \sigma} \bm h_\Omega$. 
 Then combining with Eq.\ref{Eq:lhs} we obtain the following equation 
 with collision integrals $\hat {\cal I}_{o}$ and 
 $\hat {\cal I}_{so}$
 \begin{align} \label{Eq:Correction_gh0}
 & s_1\hat g_0(1) \hat g_h - 
 s_2\hat g_h \hat g_0(2) - i(\bm v_F \bm q) \hat g_h 
 \\ \nonumber
 & = 
 -  i [  \hat g_0(1) 
 \hat{\bm h}_{\Omega}\hat\tau_3  
 - 
 \hat{\bm h}_{\Omega}\hat\tau_3 
 \hat g_0(2) ]
 +  
 \hat {\cal I}_{so} +  \hat {\cal I}_{o} 
 \end{align} 
 \begin{align}   \label{Eq:CO_o}
 & \hat {\cal I}_o = [ 
 \hat g_0(1) 
\langle \hat g_h \rangle
 + 
 \langle \hat g_h \rangle \hat g_0(2)
 - 
 \\ \nonumber
 & \langle \hat g_h \rangle 
 \hat g_0(2) -  \hat g_0(1)  \hat g_h 
  ]/2\tau_{o}  
  \end{align}   
  
  \begin{align}  \label{Eq:CO_so}
 & \hat {\cal I}_{so} =  
 [ 
 \hat g_0(1) 
 \bm\sigma\langle \hat g_h \rangle\bm\sigma
 + 
 3 \langle \hat g_h \rangle \hat g_0(2)
 - 
 \\ \nonumber
 & \bm\sigma \langle \hat g_h \rangle \bm\sigma 
 \hat g_0(2) - 3 \hat g_0(1) \langle \hat g_h \rangle  
  ]/6\tau_{so} 
  \end{align}
    }
    This Eq.(\ref{Eq:Correction_gh0}) coincides with the Eilenberger Eq. (\ref{Eq:Eilenberger}) expanded for the first-order correction 
    $\hat g_h$. This proves that the time-dependent spin response in metals
    is captured by the Eilenberger equation with corresponding collision integrals.  
  
  
 \subsection{Susceptibility of the spatially homogeneous system }    

First, we consider the spatially homogeneous system when the Zeeman field depends only on time and not on the spatial coordinate so that $\bm q =0$.
 The spatial dispersion of susceptibility is discussed in in the diffusive limit in Sec.\ref{Sec:SpatialDispersion}. 
In the homogeneous case the ordinary scattering drops out from Eq.\ref{Eq:Correction_gh0} since $\hat {\cal I}_o=0$. Then Eq.\ref{Eq:Correction_gh0} can be solved analytically yielding 
the frequency-resolved susceptibility  $\chi (12) = (\delta/\delta \bm h) {\rm Tr} [\bm\sigma \hat g_h (12)] $ 
 \begin{align} \label{Eq:chi_hk0}
 \chi (12) =   
 \frac{  \Delta^2 +s_1s_2 - \omega_1\omega_2}
 {s_1s_2 (s_1+s_2 + 4/3\tau_{so})} ,
 \end{align}
 where  $\omega$ are fermionic Matsubara frequencies,
 $\omega_1=\omega - \Omega$, $\omega_2=\omega$, 
 $s_{1,2}=\sqrt{\omega_{1,2}^2+\Delta^2}$. 
Substituting this expression to the  analytical continuation rule (\ref{Eq:SpinAnalyticalContinuation})
we obtain the frequency dependent spin  susceptibility 
$\chi_h = \chi_h(\Omega)$.    
It is interesting to note that this response function (\ref{Eq:chi_hk0}) is identical to that which 
determines the finite-frequency conductivity of a superconductor.

We can obtain analytical results in several important limiting cases.
For the {\bf (i) normal metal}  $\Delta=0$
  Eqs.(\ref{Eq:chi_hk0},\ref{Eq:SpinAnalyticalContinuation}) yield (see detailed calculation in Appendix Sec.\ref{AppSec:NormalMetal})
 \begin{align}  \label{Eq:NormalMetalSusc}
 \chi_h (\Omega) = \frac{2( 2 /3\tau_{so} + \Gamma)}
 { 2( 2 /3\tau_{so} + \Gamma)  -i \Omega}
 \end{align}
 In this case the only contribution is provided by the first term in Eq.\ref{Eq:SpinAnalyticalContinuation}.
As one can in the absence of spin relaxation $\Omega\tau_{so} \to \infty$ and $\Gamma \to 0$ the susceptibility
is vanishes. Physically this result is quite transparent because without
 relaxation the spin projection on the oscillating Zeeman field remains a good quantum number.  Let us check that this result remains valid in the superconducting state.
 For that we consider the limit of  {\bf (ii) superconductor without spin relaxation}. In this case
using following relations
  $s_1^2 - s_2^2 = \omega_1^2 - \omega_2^2$
  and $2(\omega_1\omega_2- \Delta^2 -s_1s_2) = 
  (\omega_1+ \omega_2)^2 - 
  (s_1 + s_2)^2 $  Eq.\ref{Eq:chi_hk0} can be simplified  as follows, 
  see details in Appendix \ref{Sec:AppNoRel} 
   \begin{align}
   \chi(12) =  \frac{2}{\Omega}
    \left( 
    \frac{\omega_1}{s_1} 
    -
    \frac{\omega_2}{s_2}    \right)
   \end{align}
  Thus making the analytical continuation and neglecting terms of the order $\Gamma/\Omega$ we obtain 
 \begin{align}
 \chi_h (\Omega) =  -1 - \int_{-\infty}^\infty \frac{d\varepsilon}{2\Omega}
 [ N(\varepsilon_1)n_0(\varepsilon_1) -
 N(\varepsilon_2) n_0(\varepsilon_2) ]  
 \end{align}  
 where $N(\varepsilon)$ is the normalized DOS, $\varepsilon_1 = \varepsilon - \Omega$ and $\varepsilon_2 = \varepsilon$.  
 One can see that this expression yields $\chi_h(\Omega)=0$
 irrespective of the particular energy dependence of DOS. 
 This result can be qualitatively explained by the fact that in the absence of spin relaxation spin projection on the oscillating Zeeman field axis is a conserved quantity.  
 
 Form this limiting case one can clearly see that to obtain the correct result it is necessary to take into account all parts in the Eq.\ref{Eq:SpinAnalyticalContinuation}. Indeed, 
 the contribution of the first term in Eq.\ref{Eq:SpinAnalyticalContinuation} is proportional to 
 $\int d\varepsilon  [\hat g_0^R(1) - \hat g_0^A(2) ] \partial_\varepsilon n_0 \approx 2 \Omega/\Delta$ at low temperatures. 
 This contribution is cancelled by the second term in Eq.\ref{Eq:AnalyticalContinuationGen} to yield $\chi_h(\Omega)=0$ for 
 $\tau_{so}^{-1}=0$. 

  As we have obtained in the normal metal limit, the contribution of the first term in spin susceptibility (\ref{Eq:SpinAnalyticalContinuation}) is of the order $\Omega\tau_s$ for weak spin relaxation $\Omega\tau_s \gg 1$. Thus 
 when   $\tau_{so}\Delta \gg 1$ 
  the contribution of second term can be neglected.   For stronger spin-orbit relaxation such an approximation which has been used in as it has been done in previous works\cite{inoue2017spin, taira2018spin} is inaccurate. Below we confirm this conclusion by evaluation Eq.\ref{Eq:SpinAnalyticalContinuation} numerically. 
 
 
  
  Let us now considered the opposite limit of {\bf (iii) superconductor with strong spin relaxation $\tau_{so}\Delta \ll 1 $} and small frequencies 
  $\Omega\ll \Delta$. In this case from the general Eq.\ref{Eq:chi_hk0} we obtain 
 \begin{align} \label{Eq:chi_hk0_strong_rel}
 \chi (12) =   
 \frac{3\tau_{so}}{4}
 \left( \frac{  \Delta^2  - \omega_1\omega_2}
 {s_1s_2 } +1\right) ,
 \end{align}
 Substituting this expression into the analytical continuation 
 rule (\ref{Eq:SpinAnalyticalContinuation}) after some algebra we get  
 \begin{align} \label{Eq:chi_hk0StrongSOExpansion1}
 & \frac{4}{3\tau_{so}}  \frac{{\rm Im} \chi_h}{\Omega} 
 =   
 \int_{-\infty}^{\infty} 
 d\varepsilon
 \left( \Delta^2/\varepsilon^2 +1 \right) N^2
  \partial_\varepsilon n_0
   \end{align}     
  From this expression one can see analytically that the dissipative part of the susceptibility vanishes in the zero-temperature limit. 
 
}

 

  \begin{figure*}
  \centerline{ 
  $ \begin{array}{c}
  \includegraphics[width=0.24\linewidth]
  {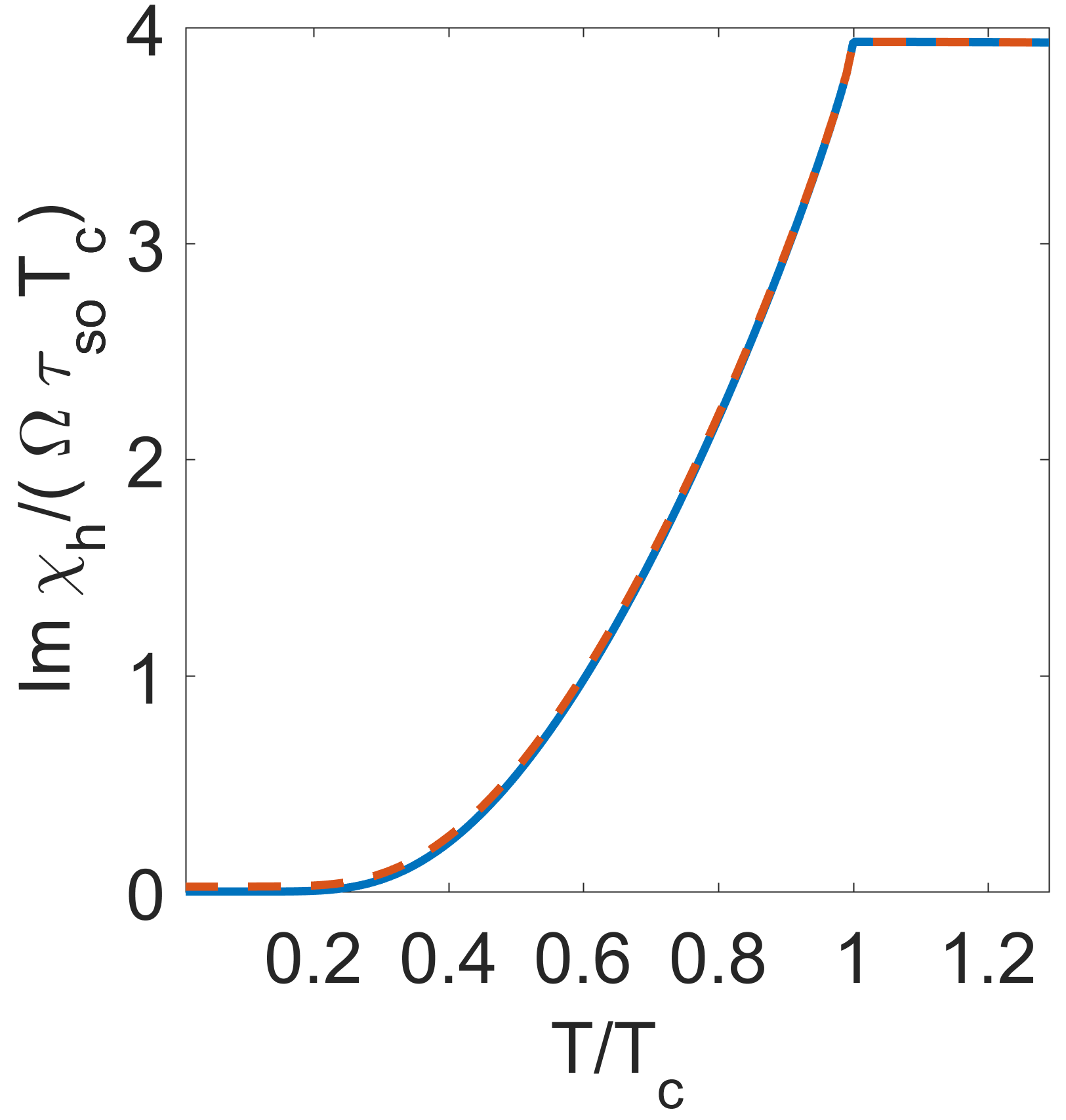}
  \put (-90,135) 
  {\large {\color{black} {\bf (a)}
  $\bm{ \tau_{so}T_c=100}$}}
  \;\;
   \includegraphics[width=0.215\linewidth]
 {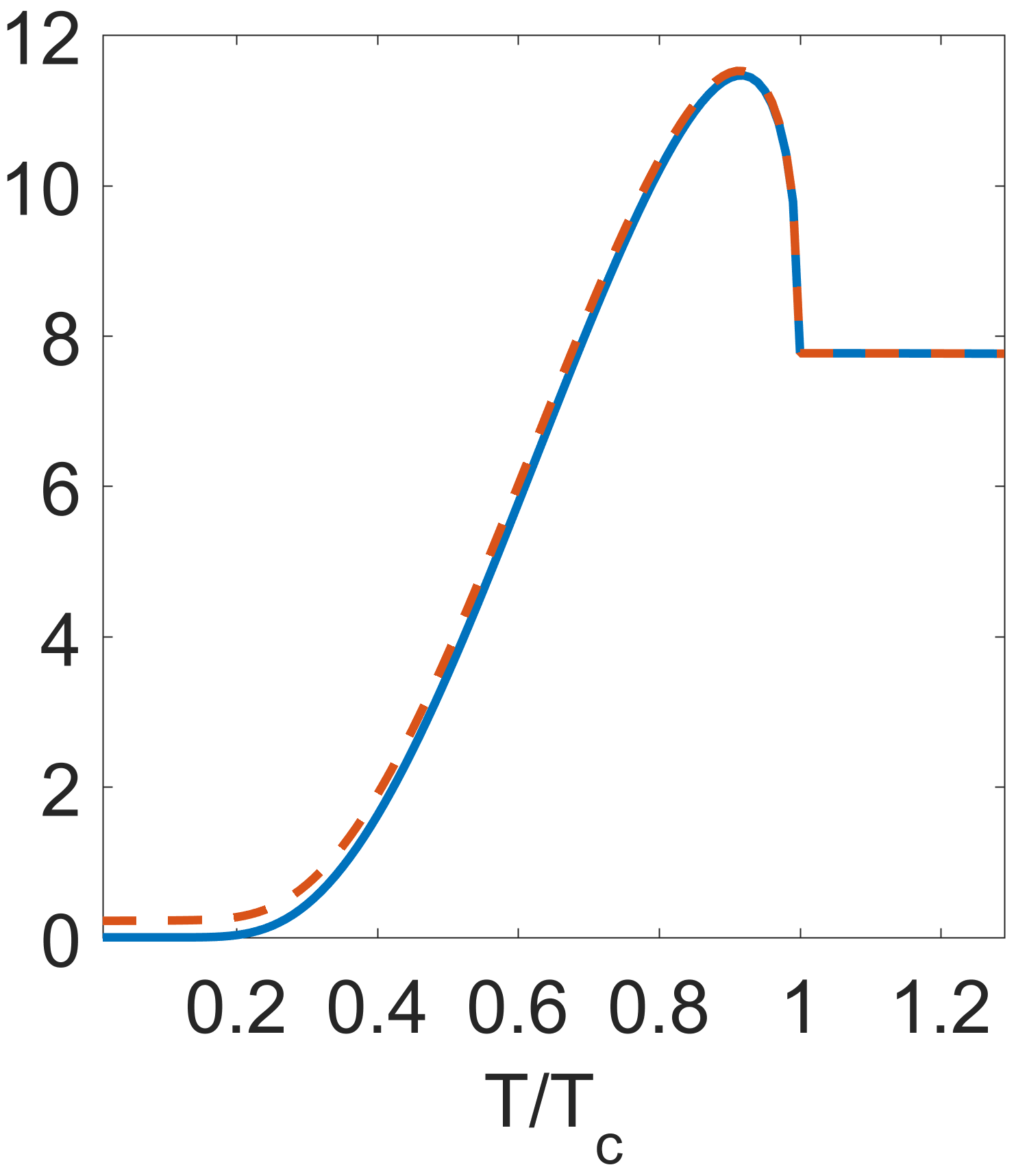}
  \put (-90,135) {\large {\color{black} {\bf (b)}
   $\bm{ \tau_{so}T_c=10}$}}
  \;\;
  \includegraphics[width=0.22\linewidth]
 {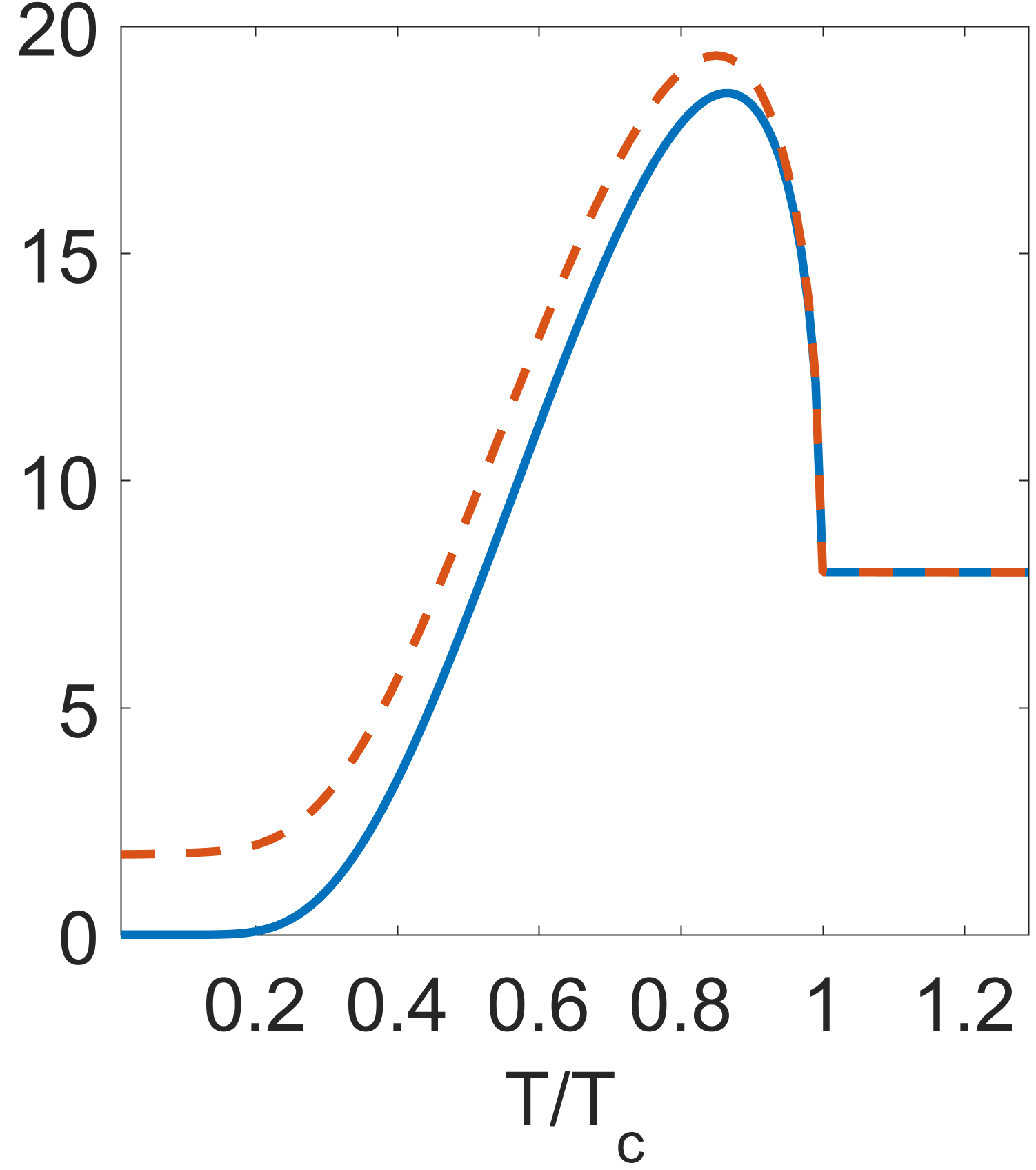} 
 \put (-90,135) {\large{\color{black} \bf (c)
  $\bm{\tau_{so}T_c=1}$ }}
  \;\;
  \includegraphics[width=0.22\linewidth]
 {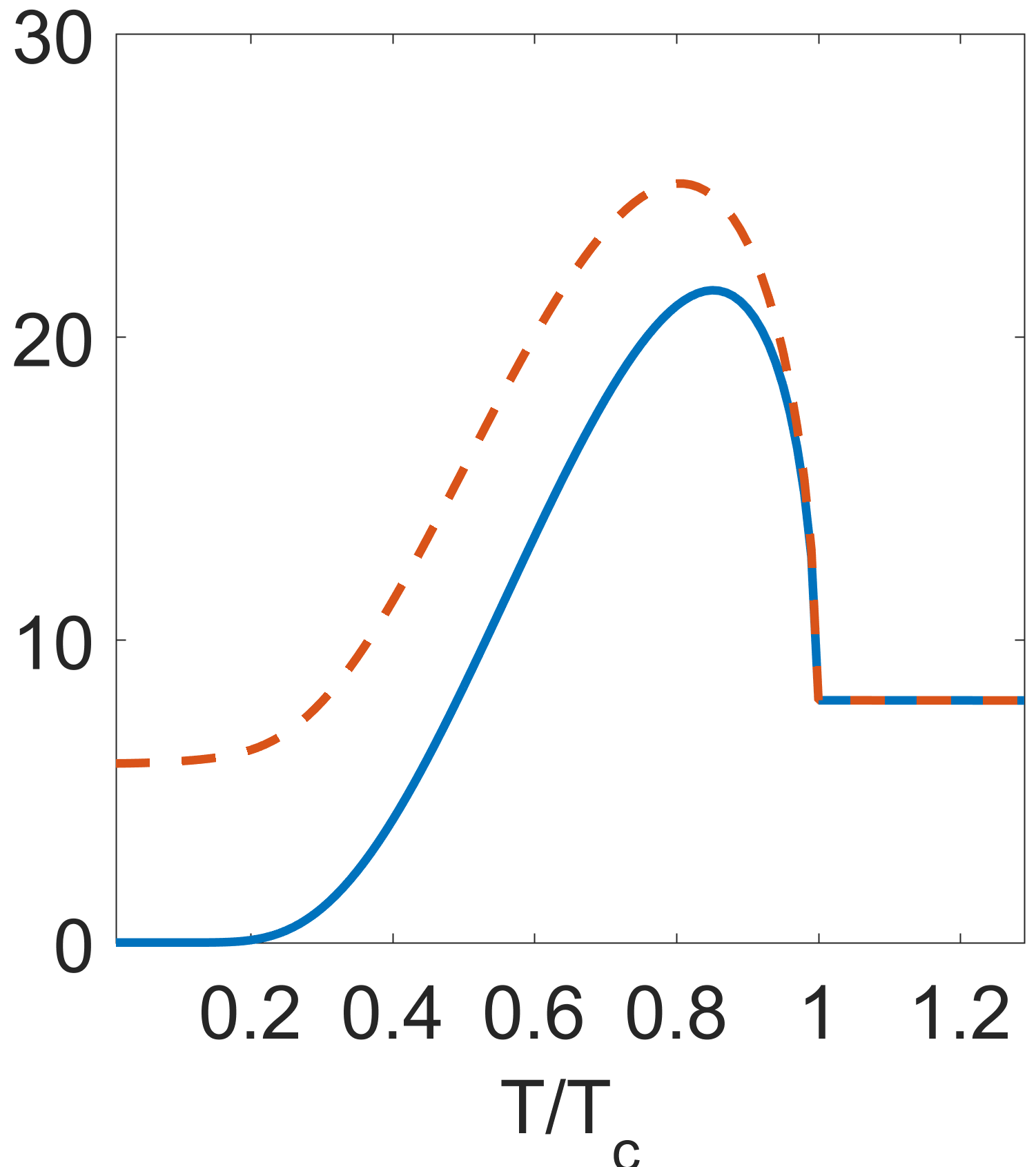} 
 \put (-90,135) {\large {\color{black} \bf (d) 
 $\bm{\tau_{so}T_c=0.1}$ }}
  \;\;
  \\
  \includegraphics[width=0.12\linewidth] {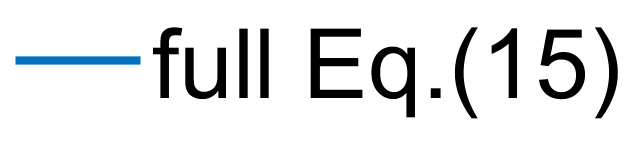} 
  \;\;\;
   \includegraphics[width=0.20\linewidth] {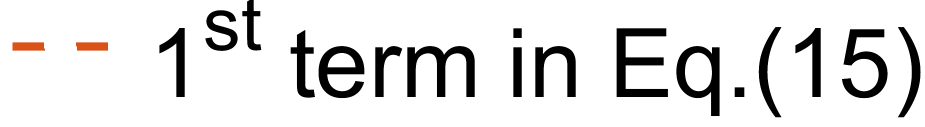} 
 \end{array}$
 }
  \caption{\label{Fig:2}  
  Comparison of the contributions to the dissipative spin response ${\rm Im} \chi_h$ given by the both terms in Eq.\ref{Eq:SpinAnalyticalContinuation} (solid blue lines) and only the first term in  Eq.\ref{Eq:SpinAnalyticalContinuation} (red dashed lines). The parameters are 
   $\Gamma = 0.001 T_c$, $\Omega =0.01 T_c$ and spin-orbit scattering time  $\tau_{so} T_c$ is 
   (a)$100$, (b) $10$, (c) $1$, (d) $0.1$. 
   }
    \end{figure*}

\begin{figure*}
 \centerline{
 $ \begin{array}{c}
 \includegraphics[width=0.24\linewidth]
 {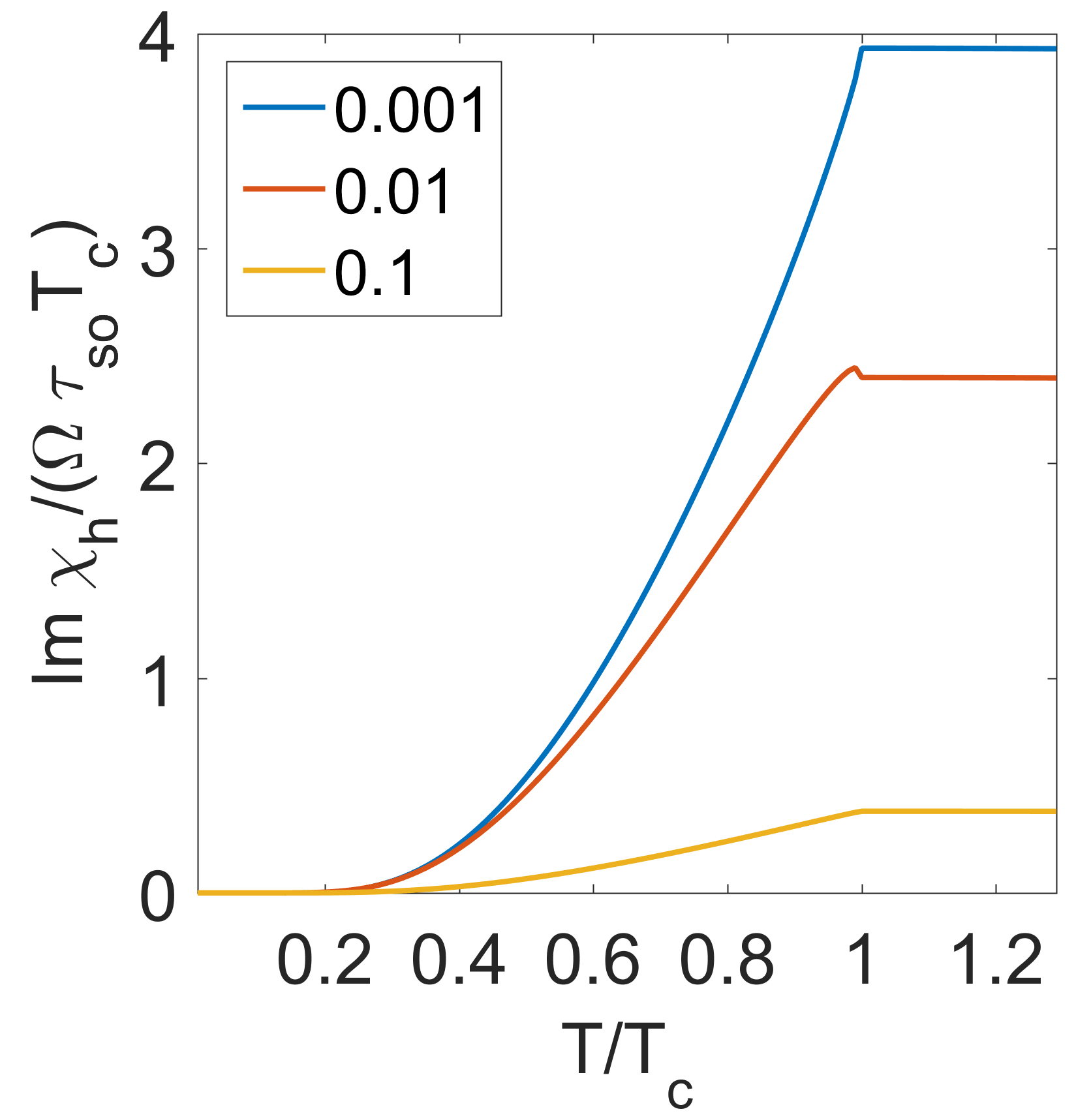}
 \put (-90,135) 
 {\large {\color{black} {\bf (a)} 
 $\bm{ \tau_{so}T_c=100}$}}
   \put (-63,113) { {\color{black}  $\Gamma/T_c$}}
 \;\;
   \includegraphics[width=0.215\linewidth]
 {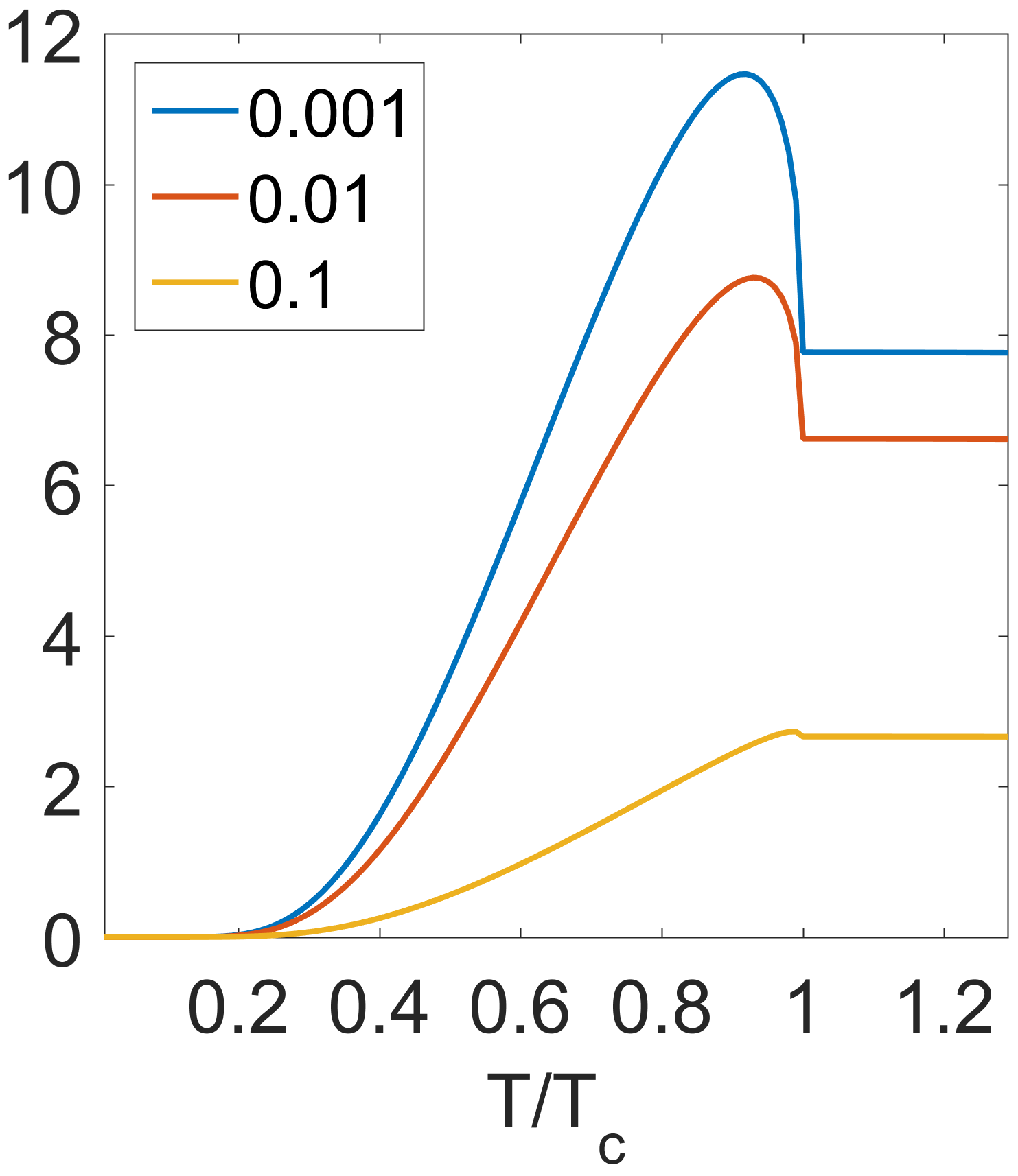}
  \put (-90,135) {\large {\color{black} {\bf (b)} 
  $\bm{ \tau_{so}T_c=10}$}}
  \;\;
 \includegraphics[width=0.22\linewidth]
 {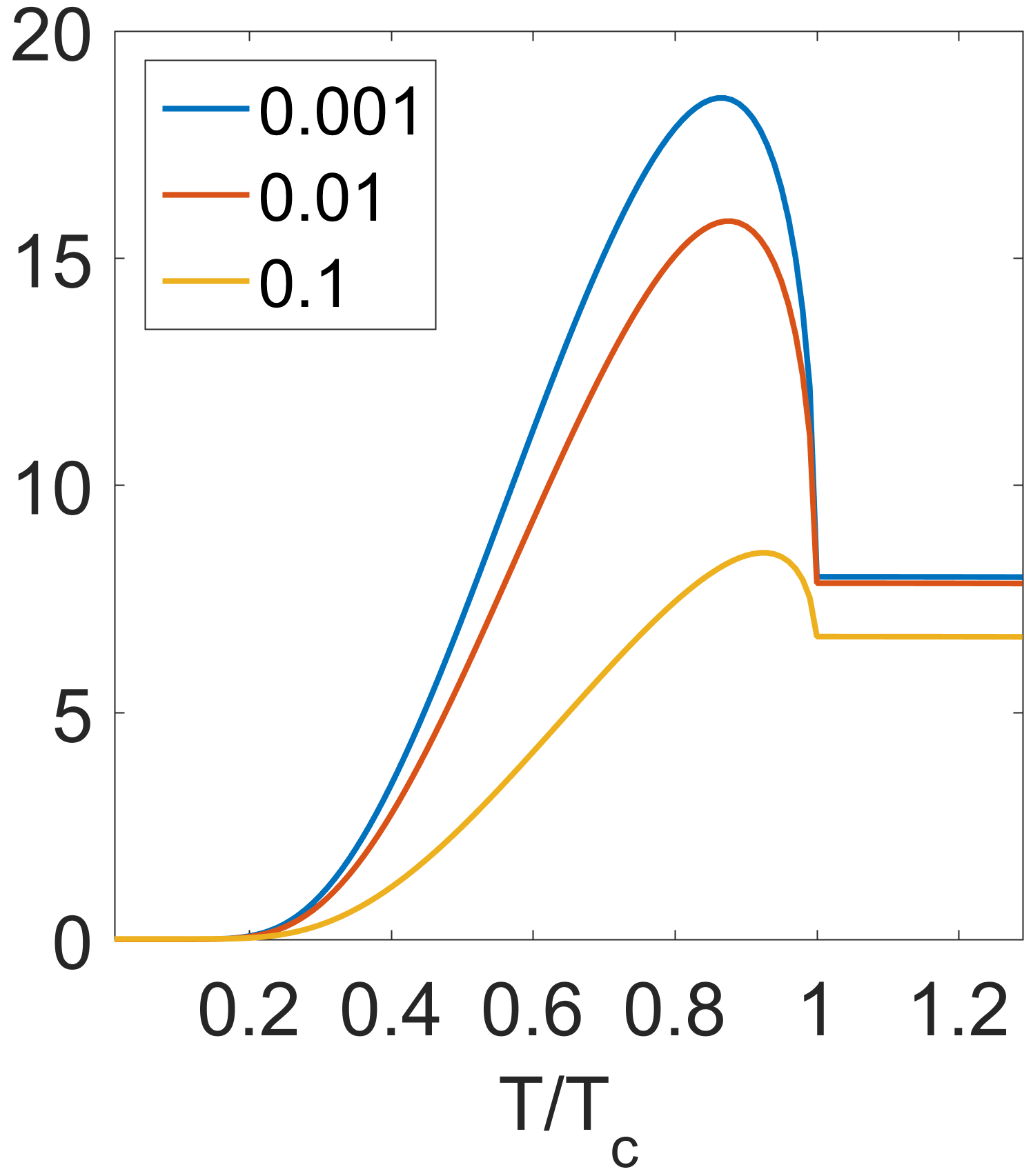} 
 \put (-90,135) {\large{\color{black} \bf (c) 
 $\bm{\tau_{so}T_c=1}$ }}
  \;\;
  \includegraphics[width=0.22\linewidth]
 {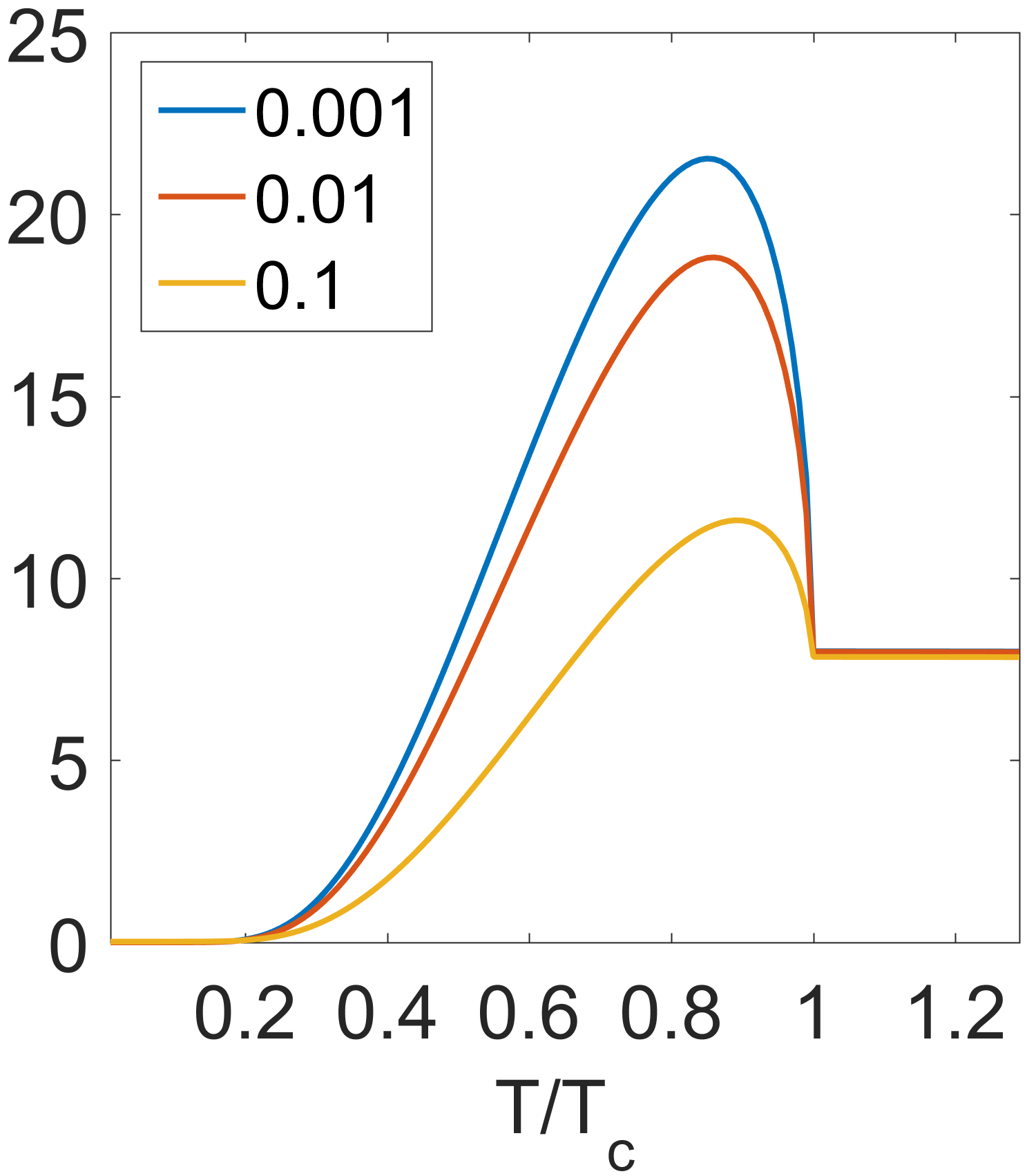} 
 \put (-90,135) {\large {\color{black} \bf (d) 
 $\bm{\tau_{so}T_c=0.1}$ }}
  \;\;
 \end{array}$
  }
  \caption{\label{Fig:3}    
  Temperature dependencies of the dissipative part of spin susceptibility ${\rm Im} \chi_h$ at small frequency  $\Omega =0.01 T_c$. In each panel curves from top to bottom correspond to the  Dynes parameter values $\Gamma/T_c = 0.0001;\; 0.01; \; 0.1$. 
The  spin-orbit scattering time $\tau_{so} T_c$ is 
   (a)$100$, (b) $10$, (c) $1$, (d) $0.1$. 
   }
    \end{figure*}

 
 {\bf General case. }
  Now let us consider the behaviour of spin susceptibility in the wide range of parameters by evaluating numerically the  integral in Eq.\ref{Eq:SpinAnalyticalContinuation}.
  First, we compare the results given by the full  Eq.\ref{Eq:SpinAnalyticalContinuation} with the contribution of only the first term. The sequence of plots in Fig.\ref{Fig:2} show temperature dependence of the dissipative part ${\rm Im} \chi_h$ at $\Omega =0.01 T_c$, Dynes parameter $\Gamma =0.0001T_c$ and several values of the spin-orbit scattering rate. The dependencies given by the full Eq.\ref{Eq:SpinAnalyticalContinuation} are shown by the blue solid curves while the dependencies given only by the first term in Eq.\ref{Eq:SpinAnalyticalContinuation} are shown by the red dashed curves.   
  One can see that for weak spin-orbit scattering $\tau_{so}T_c \gg 1$ these curves  coincide, according to the conclusion we have made based on the analysis of limiting cases above. 
  However, there is a large discrepancy for stronger spin-orbit relaxation $\tau_{so} T_c <1$. 
  Note that the behaviour of dashed curves is similar to that which  has been obtained for the dissipation signal in previous works\cite{inoue2017spin}. That is, at $\tau_{so}T_c <1$ they significantly deviate from zero at $T\to 0$.  
  As we have noted, the finite value of ${\rm Im} \chi_h$ at in the low-temperature limit is physically incorrect. 
  On the other hand, the solid curves  always demonstrate the correct behaviour going to zero in the limit $T\to 0$. Thus, the numerical analysis also confirms that both terms in the Eq.\ref{Eq:SpinAnalyticalContinuation} contribute to the dissipative part of the spin response in the superconducting state. 
  
  
  Next, let us consider how the temperature dependencies of 
  ${\rm Im} \chi_h $ at $\Omega = 0.01 T_c$  change with the Dynes parameter. 
  The sequence of plots for the three values of 
  $\Gamma/T_c = 0.001; \; 0.01;\; 1$ is shown in Fig.\ref{Fig:3}
  for different values of the spin-orbit relaxation rate.   
  One feature demonstrated by these curves is that 
  the peak in the temperature dependencies becomes less pronounced and disappears for weak spin relaxation. At the same time there relative hight of the peak almost does not change between strong $\tau_{so}T_c =1$ (Fig.\ref{Fig:3}c) and very strong $\tau_{so}T_c =0.1$ (Fig.\ref{Fig:3}d) spin relaxation. 
  Besides that, one can see that the height of the peak is 
  strongly suppressed by increasing Dynes parameter. For the realistic value in the superconductor NbN $\Gamma =0.1 T_c$
  the relative hight of the peak is about $0.2 - 0.5$ of the normal metal value at $T>T_c$. This increase is by the order of magnitude  weaker than the relative peak heights of $2-3$ observed in  spin pumping experiment in GdN/NbN bilayers \cite{yao2018probe}. Therefore one can assume that there should be a different explanation of the this experiment 
  rather than the peaked behaviour of spin susceptibility\cite{inoue2017spin}.

  \begin{figure}
 \centerline{
 $ \begin{array}{c}
  \includegraphics[width=0.495\linewidth]
 {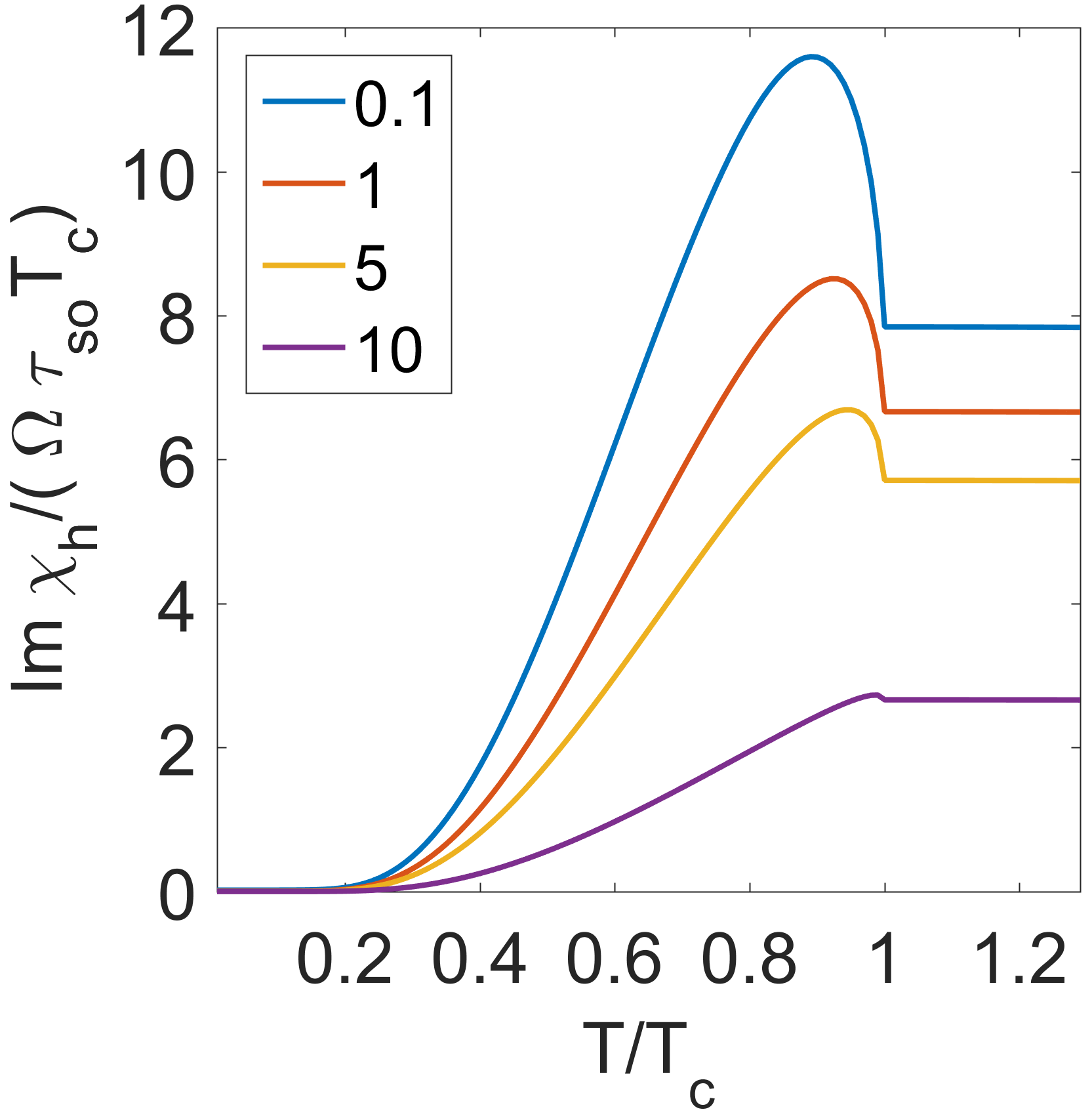}
  \put (-90,130) {\large {\color{black} \bf  (a) 
  $ \bm { \Gamma=0.1T_c }$} }
  \put (-73,113) { {\color{black}  $\tau_{so}T_c$}}
  \;\;
  \includegraphics[width=0.45\linewidth]
 {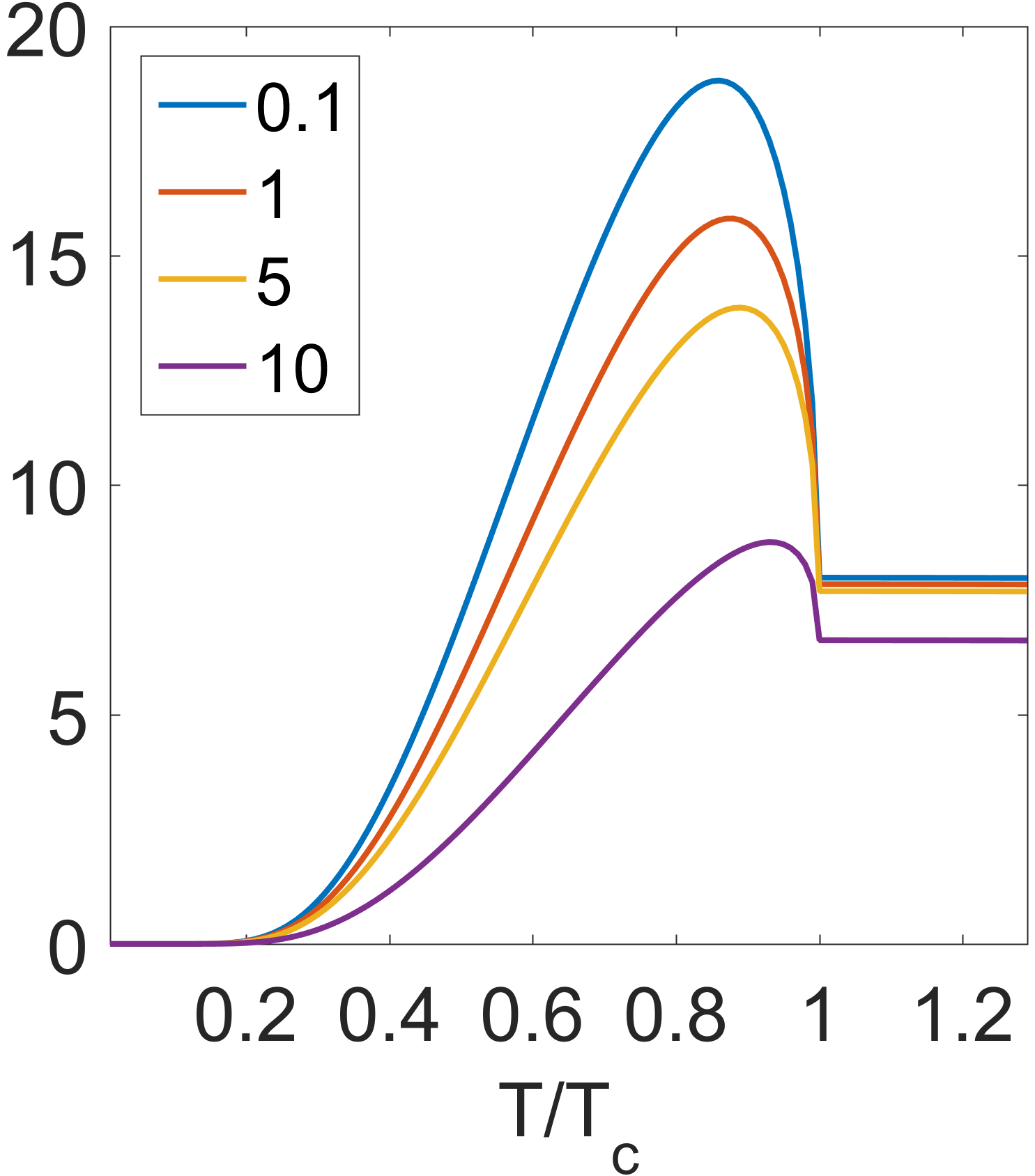} 
 \put (-90,130) 
 {\large{\color{black} \bf (b) $ \bm {\Gamma=0.01T_c }$ }}
  \end{array}$
 }
  \caption{\label{Fig:4}    
  Temperature dependencies of the dissipative part of spin susceptibility at $\Omega =0.01 T_c$ and different values of the Dynes parameter (a) $\Gamma =0.1 T_c$; (b) $\Gamma = 0.01 T_c$. Curves from top to bottom in each panel correspond to the spin-orbit scattering times $\tau_{so} T_c = 0.1; \; 1;\; 5;\; 10$.   
   }
    \end{figure}

  Now let us consider the behaviour of spin susceptibility at larger frequencies comparable with superconducting energy scales $\Omega \sim T_c$. In this case it is interesting to consider both the dissipative and the non-dissipative parts  of spin susceptibility. As we show below they are responsible  
  for the damping and field-like spin torque contributions to the spin dynamics. In Fig.\ref{Fig:5} we plot the relevant quantities 
  ${\rm Im} \chi_h (\Omega) /\Omega$ which contributes to the excess Gilbert damping and ${\rm Re} \chi_h  (\Omega) - {\rm Re} \chi_h  (0)$ which contributes to the shift of the ferromagnetic resonance central frequency. 
  First, we notice that the non-monotonic temperature dependence of the dissipative part (left panels in Fig.)
disappear at the frequencies much larger than the Dynes parameter $\Omega \gg \Gamma$. For such frequencies  ${\rm Im} \chi_h$ 
 monotonically decreases with temperature and finally disappears at $T\to 0$ provided that $\Omega<2\Delta$. For $\Omega>2\Delta$ there a non-zero  signal even at $T=0$ due to the excitation of quasiparticles across the gap.

 
 \begin{figure}
 \centerline{
 $  \begin{array}{c}
   \includegraphics[width=0.495\linewidth]
 {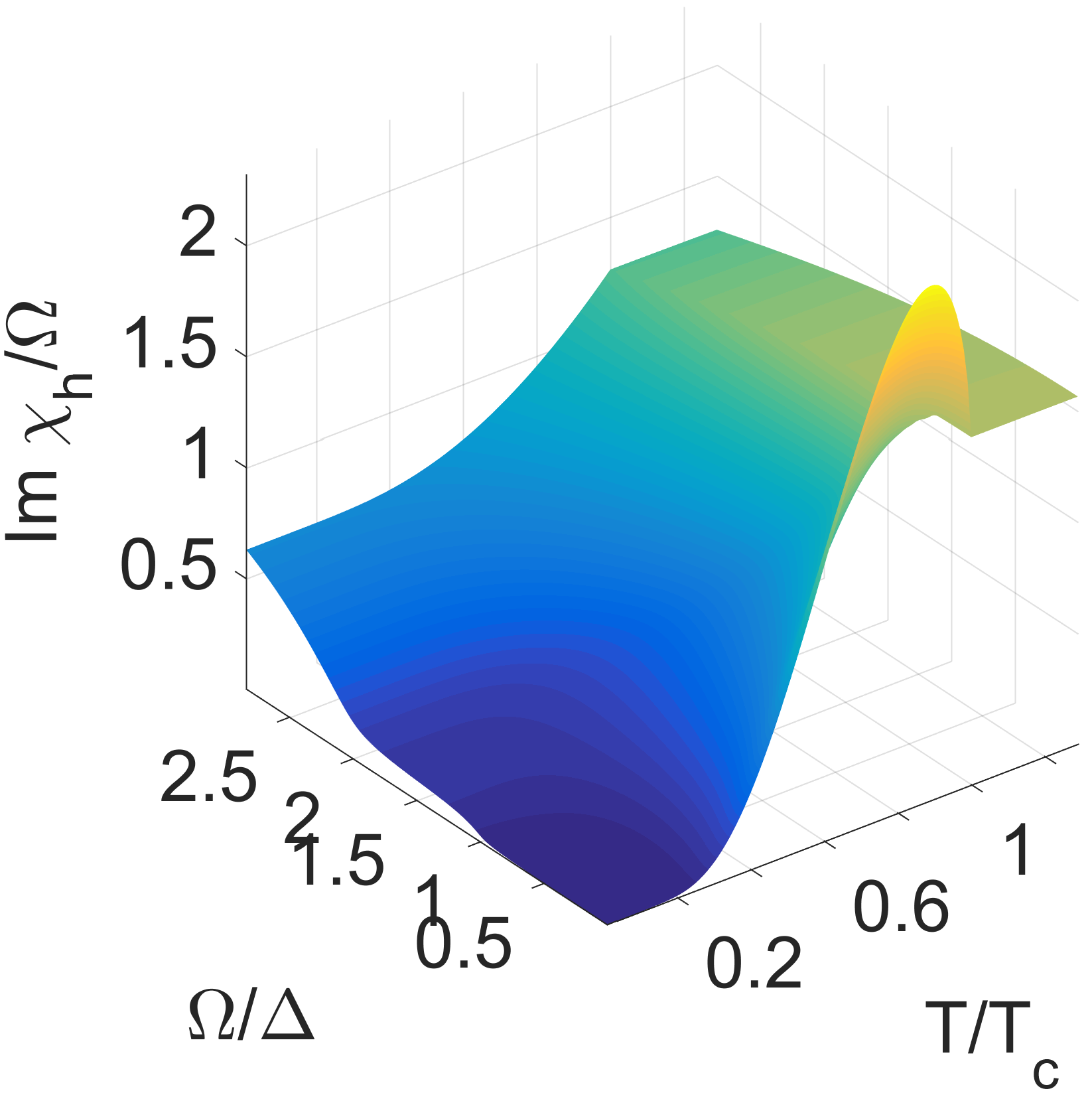}
 \;\; 
 \includegraphics[width=0.495\linewidth]
 {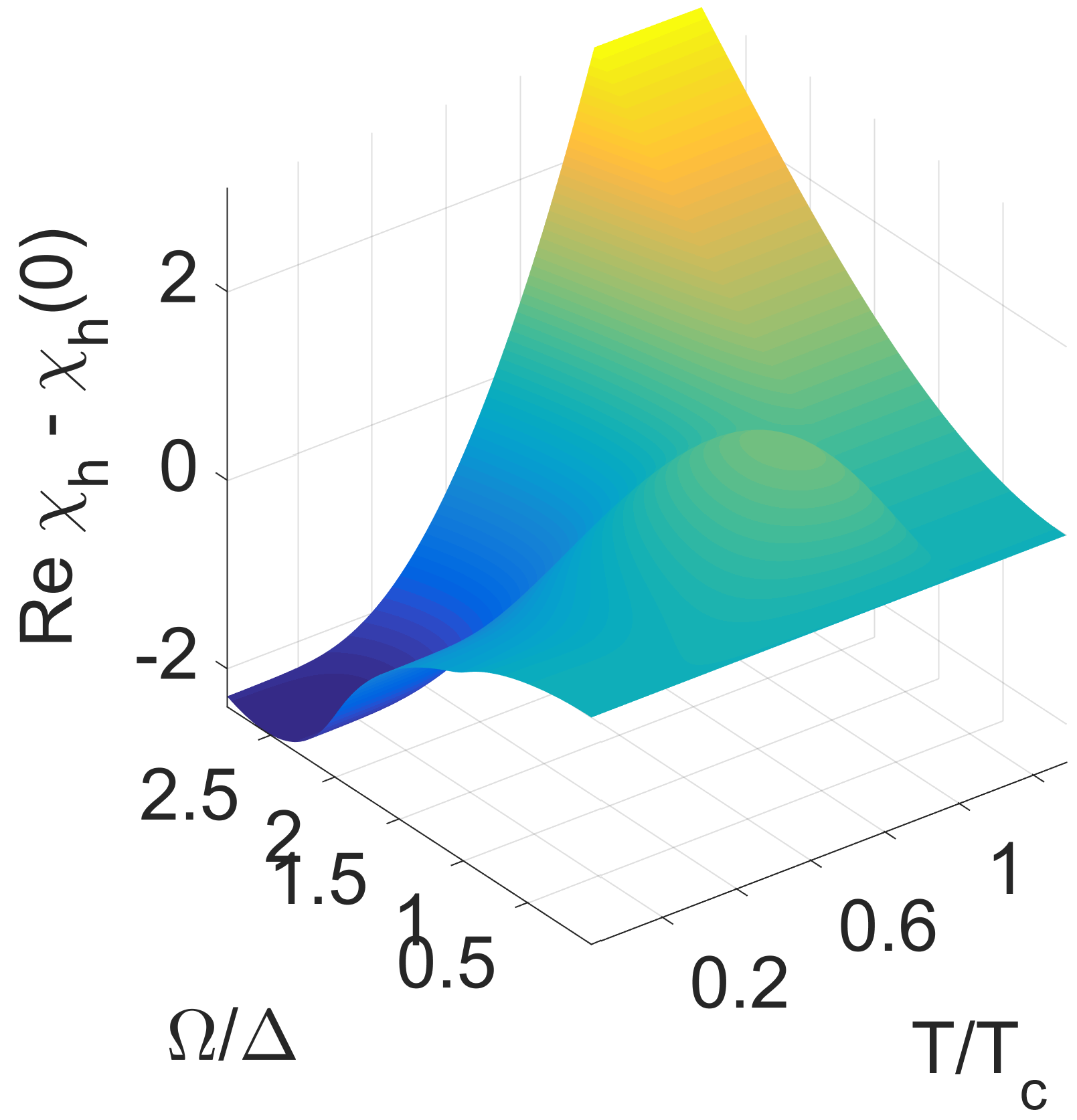} 
 \put (-150,120) {\large {\color{black} \bf (a) 
 $\bm{ \Gamma=0.1 T_c }$ }}
 \\
  \put (-150,130) {\large {\color{white} \bf (b)
   ${\bm \Gamma=0.001 t_c}$ }} 
  \includegraphics[width=0.495\linewidth]
 {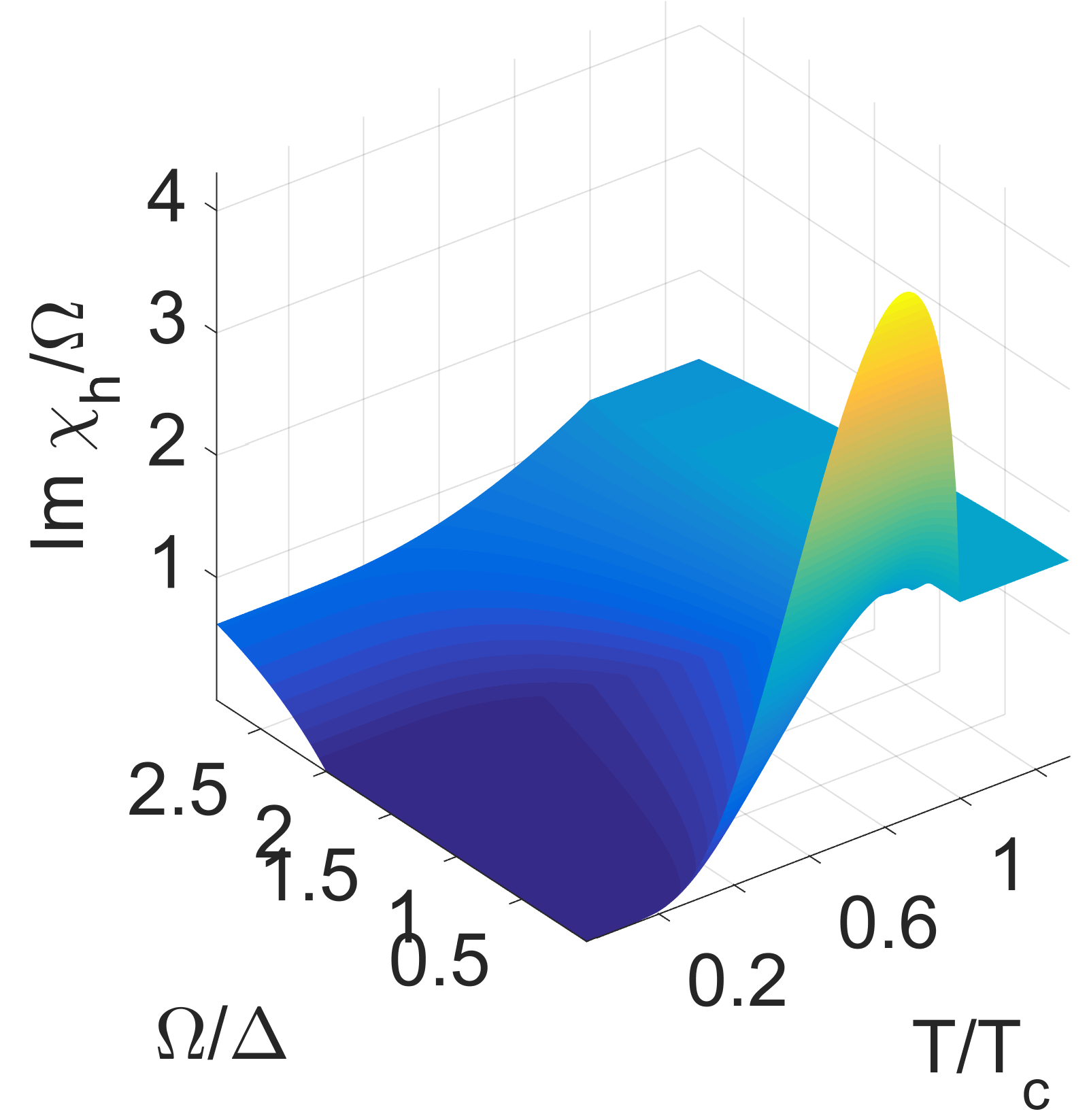} 
 \;\;  
    \includegraphics[width=0.495\linewidth]
 {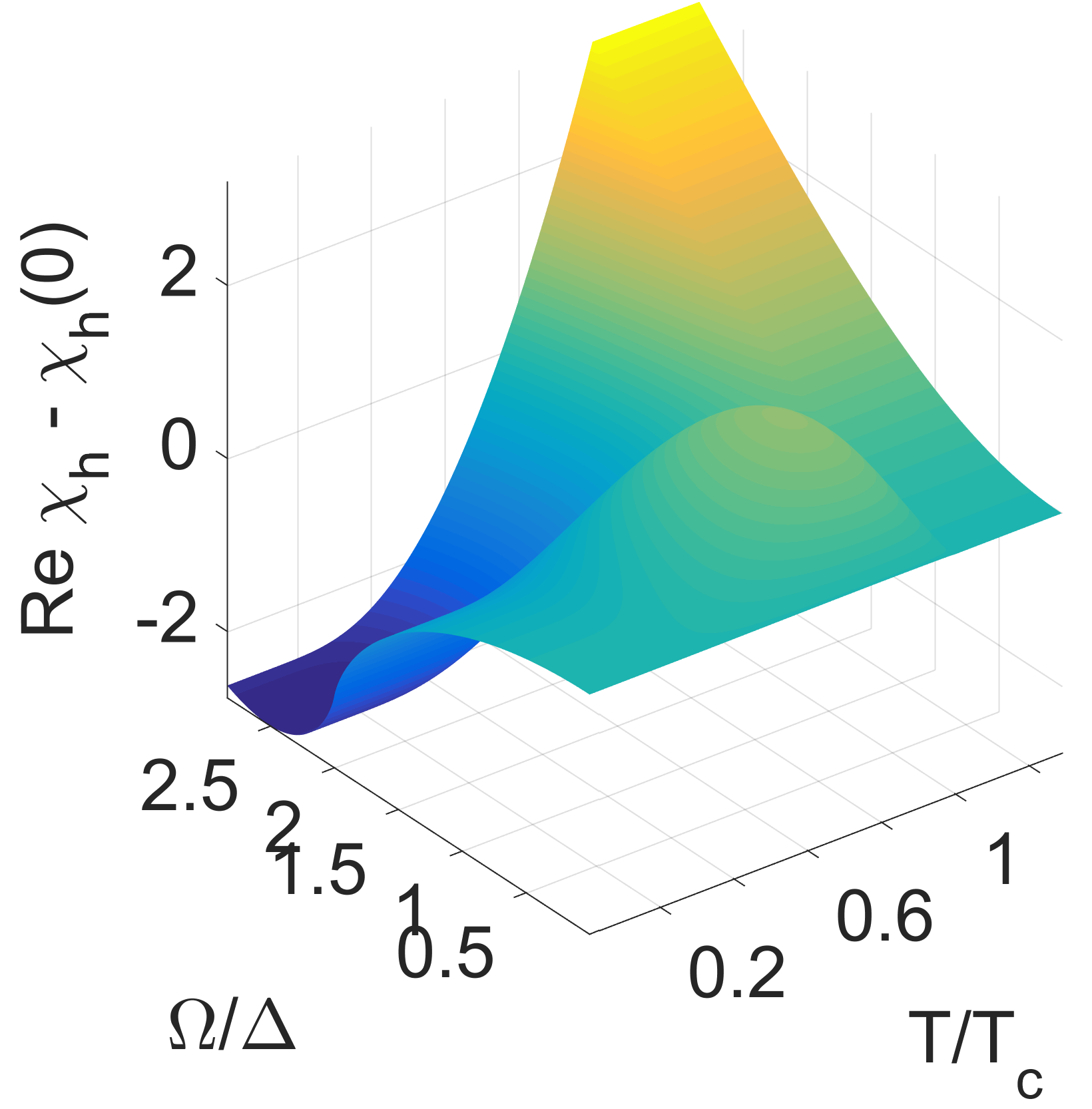} 
 \put (-150,120) {\large {\color{black} \bf (b) 
 $\bm{ \Gamma=0.001 T_c}$ }}
 \end{array}$}
  \caption{\label{Fig:5} 
  Imaginary (left row) and real (right row) parts of the spin susceptibility as functions of $T$ and $\Omega$, normalized to the zero-temperature gap $\Delta(T=0)$. The Dynes parameters are (a) $\Gamma =0.1 T_c$ and (b) $\Gamma = 0.001 T_c$. The spin-orbit scattering time is   
   $\tau_{so}T_c =1$.     
    }
 \end{figure}

\subsection{Spatial dispersion of the  susceptibility}
 \label{Sec:SpatialDispersion}
 {   
    In general, due to the presence of anisotropic term in Eq.\ref{Eq:Correction_gh0} the analytical solution is not possible for $\bm q\neq 0$. 
    However, we can still get the analytical solution in the experimentally relevant diffusive limit when the ordinary scattering rate is very large $(T_c\tau_{o})^{-1} \gg 1$. 
 In this case  Eq. \ref{Eq:Eilenberger} can be simplified by averaging over momentum directions. The isotropic part of the GF satisfies  
  Keldysh-Usadel equation 
 \begin{equation}\label{Eq:UsadelImaginary}
 -i\{\hat\tau_3\partial_t, \check g \}_t 
 +
 D\bm\nabla  ( \check g \circ \bm\nabla \check g) 
 =
 i [\hat\tau_3\hat H, \check g ]_t  
 +
 [\check \Sigma_{so}\circ, \check g]_t 
 \end{equation}
 where $D= \tau_ov_F^2/3 $ is the diffusion coefficient. 
 
  }

The spin response to the spatially-inhomogeneous Zeeman field 
$\bm h_\Omega 
 e^{i\Omega t + i qz} $
can be calculated analytically in the diffusive limit using Usadel Eq.\ref{Eq:UsadelImaginary}. Using the imaginary time representation and 
 searching the solution in the form $\hat g_h (12) e^{iqz} e^{i(\omega_1 t_1- \omega_2 t_2)}$
 we obtain the linearized Usadel equation
 \begin{align} \label{Eq:Usadel_k}
 & (s_1  + D q^2 ) \hat g_0 (1) \hat g_h - s_2 \hat g_h \hat g_0 (2) =  
 \\ \nonumber
 & i (\bm h_\Omega\bm{\hat \sigma}) 
 [\hat g_0 (1) \hat \tau_3
 - 
 \hat \tau_3 \hat g_0 (2)] 
 \end{align}
        The solution of this equation
        yields susceptibility in the form (\ref{Eq:chi_hk0}) with the 
        substitution of effective spin relaxation time 
        $4/3\tau_{so} \to 4/3\tau_{so} + D q^2$
 \begin{align} \label{Eq:chi_hk_neq0}
 \chi (12) =   
 \frac{  \Delta^2 +s_1s_2 - \omega_1\omega_2}
 {s_1s_2 (s_1+s_2 + D q^2 + 4/3\tau_{so})} 
 \end{align}
  %
  
  This expression together with Eq.\ref{Eq:SpinAnalyticalContinuation} can be used to study various phenomena related to the spin dynamics in superconductors with spin-orbit relaxation. For example, it is possible to study the effect of spin relaxation on the nuclear magnetic resonance \cite{hebel1959nuclear, masuda1962nuclear} and electron paramagnetic resonance \cite{tagirov1987spin} in superconductors. 
It is interesting that the peak in spin relaxation observed in these experiments is robust against even the very strong spin-orbit scattering as it follows from Fig.\ref{Fig:3}d and Fig.\ref{Fig:4}.   
 In the limit of weak spin relaxation there is no peak, i.e. the temperature dependence is monotonous as shown in Fig.\ref{Fig:2}a and \ref{Fig:3}a.

  \subsection{Keldysh formalism and kinetic equations}
  \label{SubSec:KinEq}
  In the general case the procedure of analytical continuation 
  is not possible and one has to consider the real time equations from the very start. 
  This brings extra complication related to the matrix structure of the contour-ordered propagator
    $\check g = 
 \begin{pmatrix} \hat g^R & \hat g^K 
 \\ 0 & \hat g^A 
 \end{pmatrix}$
 having the spectral retarded (advanced) 
 $\hat g^{R(A)}$ and the Keldysh component $\hat g^K$. 
 The matrix GF satisfies Keldysh-Usadel equation which is formally identical to the 
 Eq.\ref{Eq:Eilenberger} or \ref{Eq:UsadelImaginary} with the substitution $\partial_t \to -i\partial_t$.
  Using the normalization condition $\hat g^R\circ\hat g^K + \hat g^K\circ\hat g^A =0 $ one can introduce the parametrization of 
 the Keldysh component in terms of the distribution function  
  $\hat g^K = \hat g^R\circ\hat f -
   \hat f  \circ\hat g^A$.  
  Local spin density given by 
  \begin{align} \label{Eq:SpinCurrentOn}
  \bm S (t) = 
  -\frac{\pi\nu}{4} {\rm Tr} [ \bm{\hat \sigma} \hat\tau_3 \hat g^K(t,t)]
  \end{align}
 
 The driven state of superconductor is 
 described by the deviation of the Keldysh function from equilibrium 
  which consists of the  parts with perturbations of spectral functions $\delta \hat g^{R,A}$ and the non-equilibrium part of distribution function $\delta \hat f$. 
   In the linear response regime one can write 
   \begin{align} \label{Eq:LinearResp_gK}
  & \delta \hat g^K (12) = 
 \\   \nonumber
  &  [\hat g_0^R(1) - \hat g_0^A(2) ] \delta \hat f  + \delta g^R(12) n_0(2) - \delta g^A(12) n_0(1)    
   \end{align}
   Comparing expressions (\ref{Eq:LinearResp_gK},\ref{Eq:SpinCurrentOn}) with \ref{Eq:SpinAnalyticalContinuation}
   one can see that the first term here yields the first term in the r.h.s. of Eq. \ref{Eq:SpinAnalyticalContinuation} and $\hat f \propto n(\varepsilon_1) - n(\varepsilon_2)$. 

 In the low-frequency limit one can calculate the corrections 
 to distribution function using the kinetic equation
  \cite{RevModPhys.90.041001}
  with the driving term obtained from the gradient expansion of the 
 mixed  product in the analytical continuation of Eq.\ref{Eq:UsadelImaginary}
 \begin{align}
  [\hat H, \hat f]_t = i \hat{\bm \sigma} 
 \partial_t \bm h \partial_\varepsilon n_0 
    \end{align}
 Parametrizing the spin-dependent distribution function as $ \hat f = \hat{\bm \sigma} \bm f$ we get the kinetic equation which for the spatially homogeneous system is given by
   \begin{align}
   \label{Eq:Kin-1}
  & \partial_t\bm f + (2\Gamma + \tau_s^{-1}){\bm f} = 
   \partial_\varepsilon n_0\partial_t\bm h
    \\  \label{Eq:taus-1}
    &   \tau^{-1}_s  =
  (1/3\tau_{so})
  N^{-1} {\rm Tr} (1 - \hat g^R \hat g^A)
   \end{align} 
   where $N =  {\rm Re} {\rm Tr} [\hat\tau_3 \hat g^R ] /2$ is the normalized density of states. 
At the subgap energies $|\varepsilon| < \Delta$ 
the  spin relaxation rate (\ref{Eq:taus-1}) is not defined if the density of states is strictly zero $N=0$. However, for the finite Dynes parameter $N\propto \Gamma$ so that  $\tau_s^{-1} \propto \Gamma$. The solution of the Eq.\ref{Eq:Kin-1} yields the contribution to the spin density 
  \begin{align}
  & \chi_{kin} +1 = \frac{\Omega }{2} \int_{-\infty}^{\infty} d\varepsilon \frac{N \partial_\varepsilon n_0}{\Omega -i(2\Gamma + \tau_s^{-1}) } ,
   \end{align}
   which  coincides with the first term in Eq.(\ref{Eq:SpinAnalyticalContinuation}) in the low-frequency limit.

           
  \section{Spin pumping in superconducting films} 
  \label{Sec:SpinPumping}
   
 \begin{figure}[htb!]
 \centerline{$
 \begin{array}{c} 
 \includegraphics[width=0.7\linewidth]{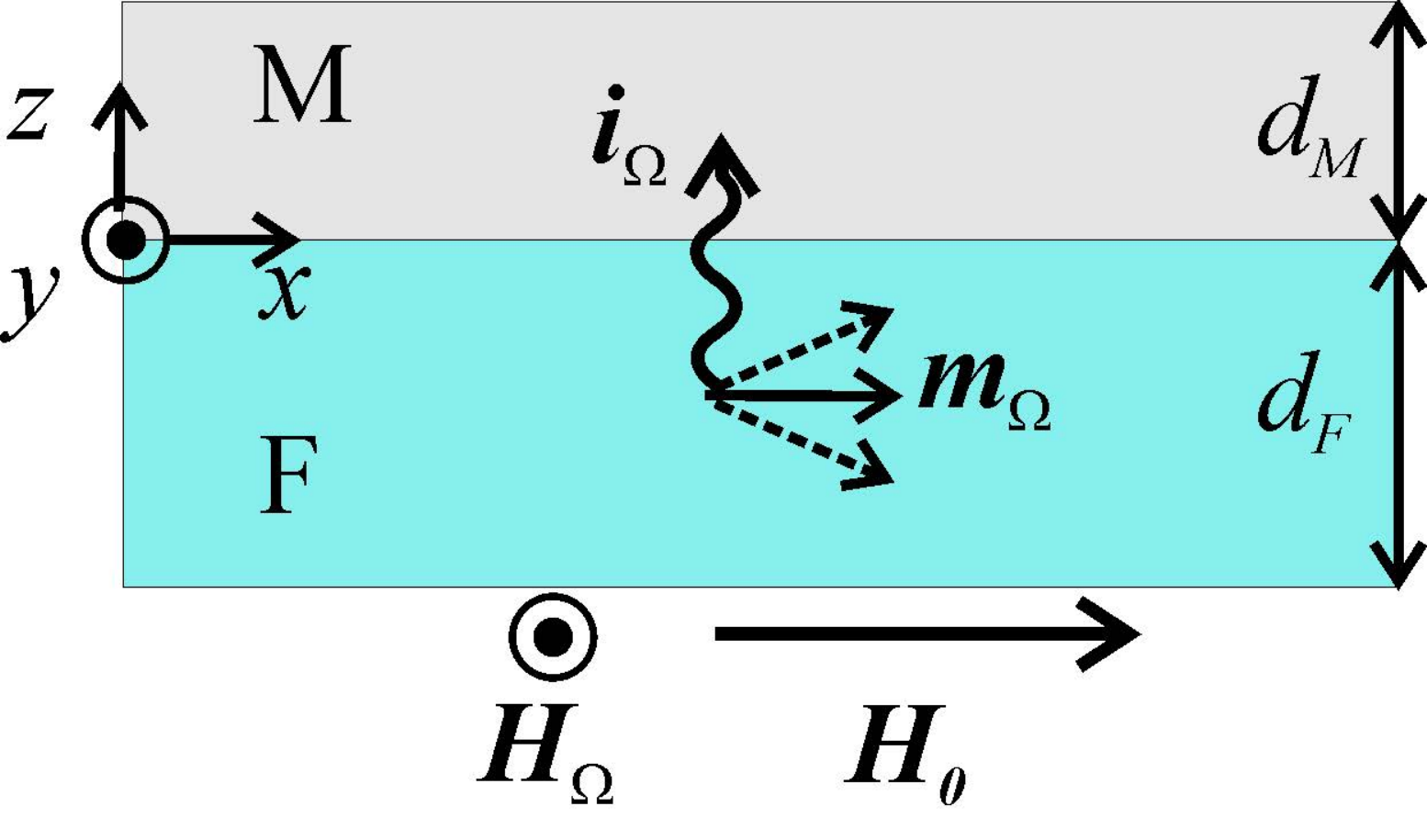} 
 \put (-90,110) {\large {\color{black} \bf (a) 
   }} 
  \\
  \includegraphics[width=0.7\linewidth]{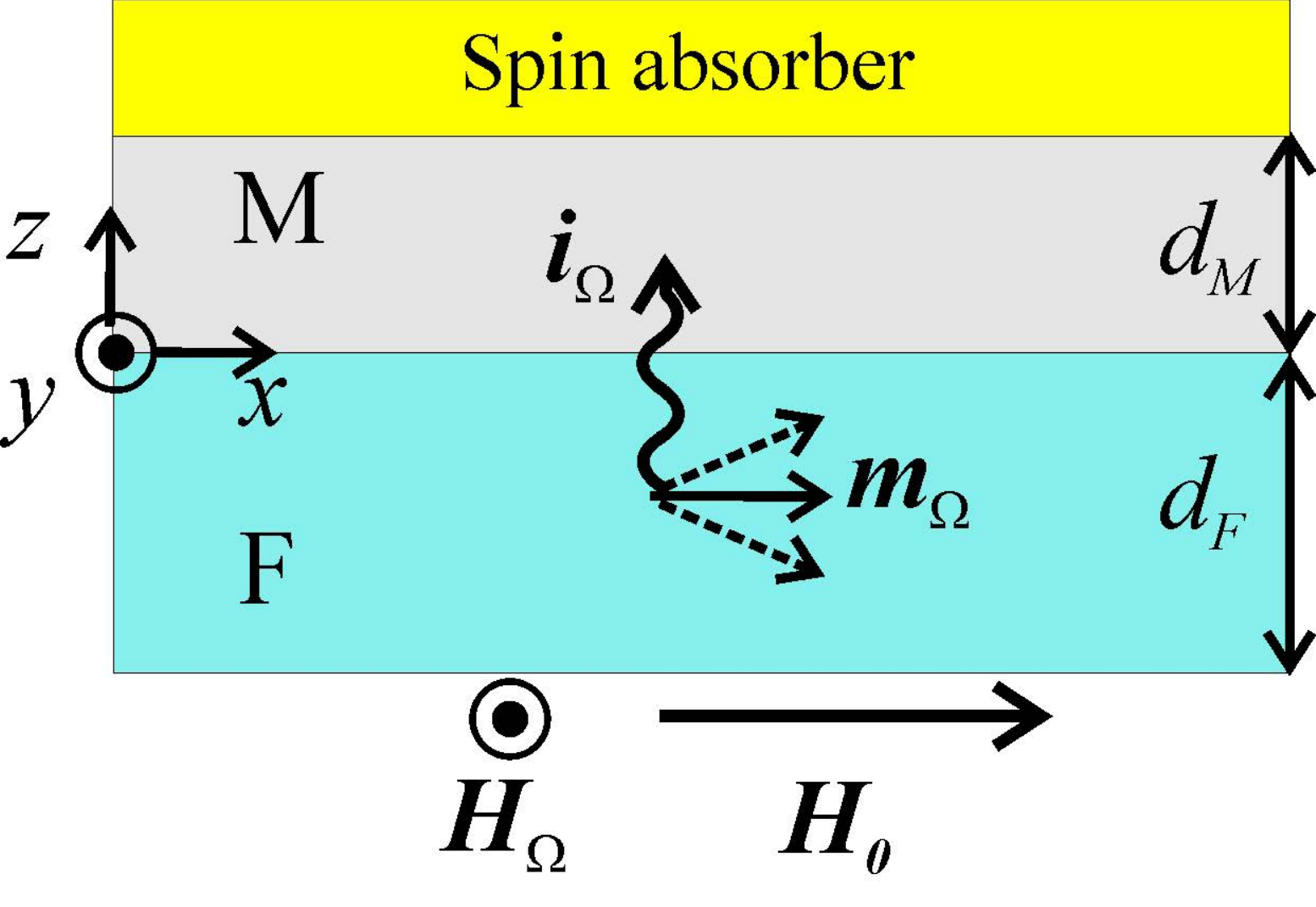} 
  \put (-90,120) {\large {\color{black} \bf (b) 
   }} 
 \end{array}$}
 \caption{\label{Fig:FMsetup} (Color online) 
  Schematic setup with the interface between metallic spin sink (M) and ferromagnetic film (F) of the widths $d_M$ and $d_F$, respectively. 
 The constant external magnetic field is $ H_0 \bm x$.  
 The magnetization precession $\bm m_\Omega e^{i\Omega t}$ is driven by the external magnetic field $ H_\Omega e^{i\Omega t} \bm y$. It generates spin current 
 $\bm i_\Omega$ pumped from F to M.  
 (a) M has interface with vacuum; (b) M has interface with the perfect spin absorber.  
   }
 \end{figure}   

   With the general expression for spin susceptibility in hand 
   we can study effects of spin pumping from the ferromagnet into the adjacent metallic film. The schematic setups are shown in Fig.\ref{Fig:FMsetup}. The metallic spin sink M has an interface with (a) vacuum and (b) perfect spin absorber. The correposnding boundary conditions are (a) vanishing spin current and (b) vanishing non-equilibrium spin polarization at $z=d_M$.
     To quantify the spin pumping effect we consider the interfacial exchange interaction between the localized spins in F and conduction elections in M. 
     Within this model the local spin polarization close to the interface $\bm S(t)$ acts as effective field for the localized magnetic moments. 
     This process can be taken into account by introducing the additional term $\bm i(t)$ into the 
     Landau-Lifshitz-Gilber equation 
 \begin{align} \label{Eq:LLG}
 & (1 + \alpha \bm m \times )\partial_t\bm m + 
 \gamma\bm m\times \bm H_{eff} =  \bm i/S_{F0}d_F 
  \\
   \label{Eq:SpinCurrentOn}
   & \bm i (t)= J_{sd}
  \bm S (t)\times \bm m (t) 
 \end{align}
 Here $S_{F0}$ is the equilibrium spin density in F, $d_F$ is the FI film thickness, $\bm H_{eff}$ is the effective field and $\alpha$
 is the intrinsic Gilbert damping coefficient.   The term $\bm i (t)$ can be interpreted as the spin current between F and M. 

    To calculate $\bm S (t)$ we use the spin susceptibility (\ref{Eq:Susceptibility0}) with the Zeeman field determined by the interfacial exchange $\bm h = J_{sd} \bm m  \delta (z)$.   
 In the linear regime the local spin polarization near F interface 
 can be written as follows
 \begin{align} \label{Eq:SOmegaChiM}
 \bm S_{\Omega} = \nu h_{eff} \chi_m \bm m_\Omega  
 \\ \label{Eq:ChiM_def}
 \chi_m (\Omega) = \sum_{n=0}^{\infty} \chi_h (q_n,\Omega) 
 \end{align}
 where we introduce the effective exchange field 
 $h_{eff} = J_{sd}/d_M $ and the local spin susceptibility $\chi_M$ which determines the response to the delta-functional Zeeman field.  
 The summation in  Eq.\ref{Eq:ChiM_def} runs over the discrete
 set of momenta given by $q_n =n \pi/d_M $ for the vacuum interface Fig.(\ref{Fig:FMsetup}a) 
 which is determined by the zero boundary condition for the spin current 
 at the interface with vacuum $z=d_M$.  
 For the strong spin sink interface Fig.\ref{Fig:FMsetup}b 
 we have $q_n =(n+1/2) \pi/d_M $ which is determined by the zero boundary  
 of the non-equilibrium spin polarization which is suppressed by the strong spin sink at $z=d_M$. 
 Derivation of this result is given in Appendix
 \ref{SecApp:CalculationGiniteThickness}.

  Taking into account the Eq.\ref{Eq:chi_hk_neq0} one can see that the only difference  introduced by the spin absorber  Fig.\ref{Fig:FMsetup}b is the modification of spin relaxation rate to $\tau_{so}^{-1} \to \tau_{so}^{-1} + D (pi/2d_M)^2 $. Therefore hereafter we will not distinguish these two cases implying that the effective spin relaxation is used.     
 
The Fourier components of the spin current (\ref{Eq:SpinCurrentOn}) is given by   
 \begin{align}
 & \bm i(\Omega)=
 \nu h_{eff}^2 d_M  [\chi_m(\Omega) - \chi_m(0)] 
 \bm m \times \bm m_\Omega  
 \end{align}

  {
  For the configuration in Fig.\ref{Fig:FMsetup} the effective field is given by 
  $\bm H_{eff}  = H_\Omega e^{i\Omega t} \bm y + B_0 \bm x$ where $B_0 = H_0 + 4\pi M$. 
   In this case the eigen frequencies of LLG Eq.\ref{Eq:LLG} satisfy the equation  
  \begin{align} \label{Eq:OmegaFMR}
 & \Omega = \sqrt{(\gamma B_0 + \delta \omega )
  (\gamma H_0 + \delta \omega )} 
  \\  
 &  \delta \omega =  i\Omega  \alpha + 
   [\chi_m(\Omega) - \chi_m(0)] T_c C 
    \end{align} 
  \begin{align}  \label{Eq:CouplingParameter}
 & C =  \frac{h_{eff}}{T_c} \frac{\nu h_{eff}}{S_{F0}}  
  \frac{d_M}{d_F}  
 \end{align} 
  
   The extra dissipation, that is the imaginary part of $\Omega$ 
   in (\ref{Eq:OmegaFMR}) can be considered resulting from the effective  Gilbert damping constant increase  
 \begin{align} \label{Eq:GDFiniteThickness}
 \delta\alpha = C  T_c {\rm Im}\chi_m/\Omega 
  \end{align} 
  In case if the film thickness is small  $d_M < min(l_{so}, \xi)$ 
  where $l_{so} = \sqrt{D\tau_{so}}$ is
   spin relaxation length and $\xi= \sqrt{D/T_c}$ is the zero-temperature coherence length ,  
  only the contribution with $n=0$ in the sum (\ref{Eq:ChiM_def}) is important. In this case the spin pumping effect is totally determined by the homogeneous spin-orbit relaxation so that $\chi_m (\Omega)\approx \chi_h(\Omega, q=0)$. 
For larger film thickness we need to take into account several terms in Eq.(\ref{Eq:ChiM_def}). 
 Only for the very large thickness $d_M \gg min(l_{so}, \xi)$ 
 the expression used in previous works \cite{ohnuma2014enhanced,inoue2017spin} is recovered in the form  
 \begin{align} \label{Eq:Gilbert}
 \delta \alpha =  \frac{\nu J_{sd}^2}{d_F S_{F0}} \int_{-\infty}^{\infty} \frac{dq}{\pi} 
  \frac{{\rm Im} \chi_h(q,\Omega)}{\Omega} .
  \end{align}

 Temperature dependencies of the   normalized excess Gilbert damping are 
 shown in Fig.\ref{Fig:5}. One can see that these dependencies are 
 qualitatively similar to that obtained in the absence of spin relaxation 
 for infinite superconducting films \cite{kato2019microscopic}. 
 They are also qualitatively similar to the temperature dependencies 
 of the NMR   
 \cite{hebel1959nuclear,masuda1962nuclear} and 
 EPR \citep{rettori1973magnetic, davidov1974electron} linewidths 
 in superconductors. 
 Note that for relatively large Dynes parameter $\Gamma = 0.1 T_c$ the peak 
 in the temperature dependencies of Gilbert damping is almost absent 
 (red curves in Fig.\ref{Fig:GD}) and superconductivity leads to the 
 monotonous suppression of the spin pumping dissipative signal. 
 This result reproduces theoretically the behaviour observed in 
 FMR experiments with Py/Nb bilayers \cite{bell2008spin}. Using large 
 Dynes parameter $\Gamma \sim T_c$ one can describe qualitatively the 
 effect of superconducting gap suppression near the surface of metallic  
 ferromagnet such as Fe or Ni. 
 At the same time the Dynes parameter $\Gamma =0.1 T_c$ corresponds to 
 the superconductors with large electron-phonon relaxation rate such as NbN.
 Therefore, provided the mechanism of spin pumping between the FI and 
 NbN superconductor is
 correctly described by the Eq.\ref{Eq:Gilbert} or 
 Eq. \ref{Eq:GDFiniteThickness} the Gilbert damping behaviour should 
 correspond to the red curves in Fig.\ref{Fig:GD} with rather weak peaks.
 The amplitude of these peaks is much smaller than has been observed in 
 the experiment \cite{yao2018probe}. This discrepancy shows the presence 
 of some other yet unknown mechanism of spin pumping which can yield 
 more pronounced peaks. The identification of such a mechanism is however 
 beyond the scope of the present paper.  
  
  \begin{figure}[h!]
 \centerline{
 $ \begin{array}{c}
 \includegraphics[width=0.525\linewidth]
 {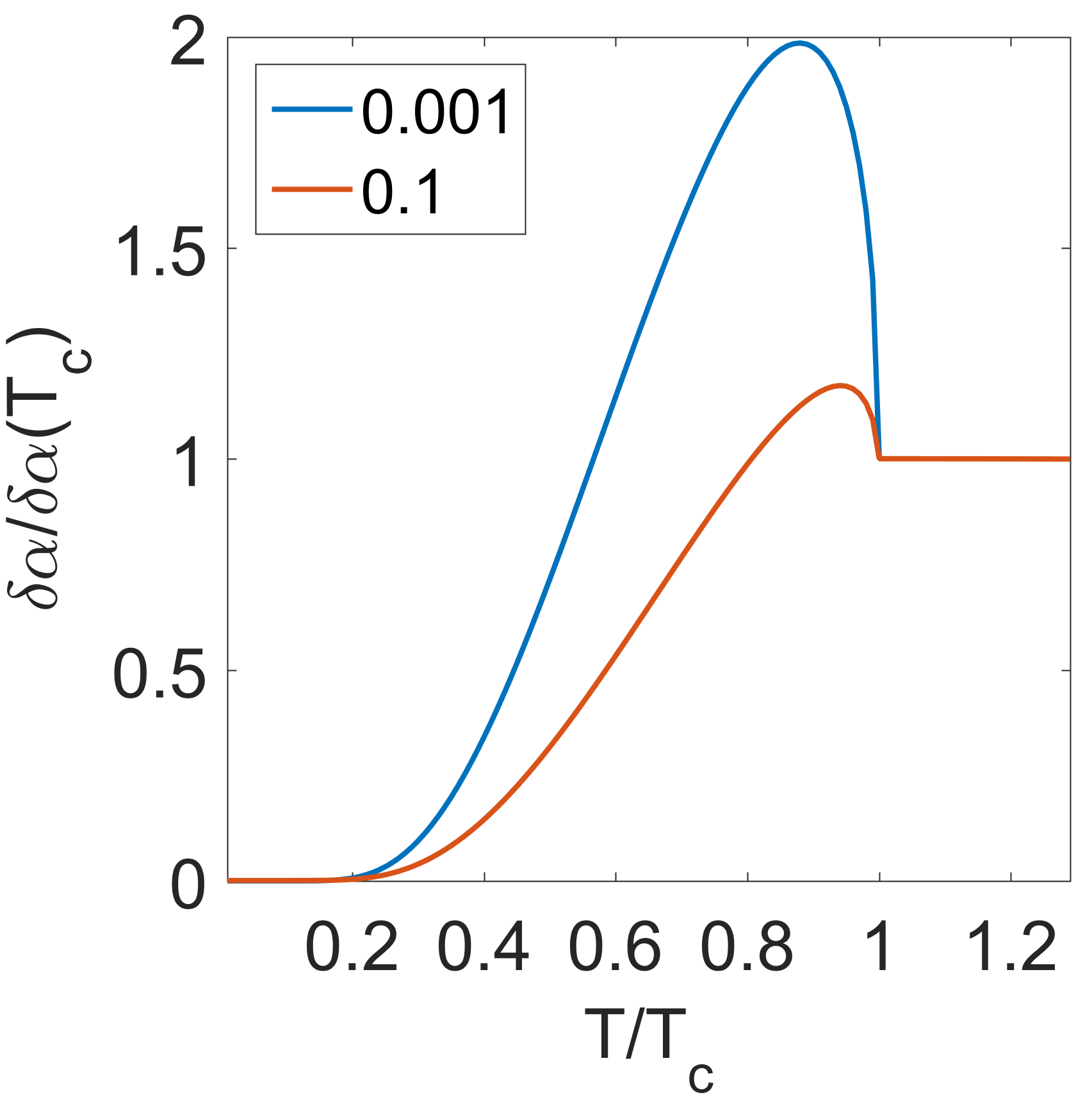}
 \put (-90,135) 
 {\large {\color{black} {\bf (a)} 
 $\bm{ d_M=3\xi }$}}
 \;\;
   \includegraphics[width=0.4775\linewidth]
 {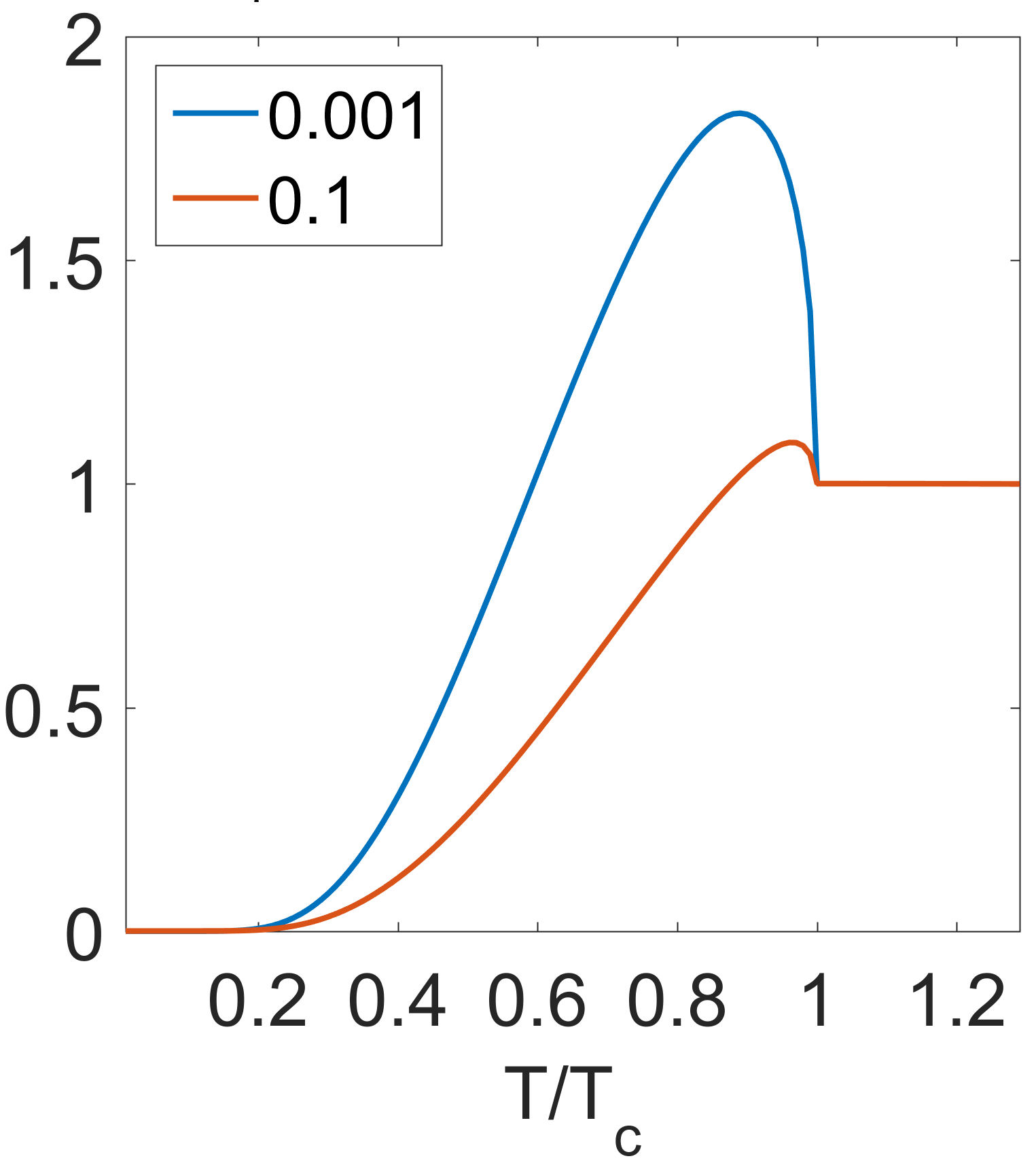}
  \put (-90,135) {\large {\color{black} {\bf (b)} 
   $ \bm {d_M=\xi/2}$}}
  \;\;
 \end{array}$
 \put (-75,50) { {\color{black}  $\Gamma/T_c$}}
  }
  \caption{\label{Fig:GD}    
  Temperature dependencies of the additional Gilbert damping coefficient $\delta \alpha$ Eq.\ref{Eq:GDFiniteThickness} at small frequency  $\Omega =0.01 T_c$. 
  In each panel curves corresponding to the Dynes parameter values $\Gamma/T_c = 0.001;\; 0.1$ are shown. 
  The spin-orbit scattering time $\tau_{so} T_c =4$ corresopnding to the normal state spin relaxation length $l_{so} = \xi/2$. 
  The metallic film thickness is (a) $d_M = 3\xi$, (b) $d_M = 0.5 \xi$. 
   }
    \end{figure}

Quite interestingly, spin relaxation and superconducting correlations lead to the 
pronounced frequency dependence of the real part of the susceptibility 
${\rm Re} \chi_h$ as shown in Fig.\ref{Fig:5}, right panels. 
 This leads to the additional contribution to the spin pumping having the form of the field-like spin torque, that is additional frequency-dependent effective field acting on the magnetization of the ferromagnet $\bm m$.
This leads to the shift of the FMR central frequency which can be 
obtained from Eq.\ref{Eq:OmegaFMR} as follows
 \begin{align}
 \delta \Omega =C T_c
 \frac{{\rm Re} [\chi_m(\Omega) - \chi_m(0)]}{2\Omega}
   \gamma (B_0 + H_0)
  \end{align}
  }
 This shift is negligible at small frequencies $\Omega T_c \ll 1$
 and $\Omega \tau_{so} \ll 1$ and small interfacial coupling between F and M films measured by the dimensionless parameter (\ref{Eq:CouplingParameter}). 
   However, it becomes significant for 
 higher frequencies and larger $C$. 
 
 To quantify the superconductivity-induced FMR frequency shift we consider the 
 system with not very strong spin relaxation $\tau_{so} T_c =1$.  
 The normalized FMR response function which according to Eqs.(\ref{Eq:LLG},\ref{Eq:OmegaFMR}) is proportional to 
  $[\Omega^2 - (\gamma B_0 + \delta \omega )
  (\gamma H_0 + \delta \omega ) ]^{-1}$. In Fig.\ref{Fig:8} we normalize this response function of its largest value at each frequency, so that it is possible to see the transformation of the FMR line as a function of temperature.  

   One can see two pronounced effects which appear with increasing the  coupling parameter. First, comparing Fig.\ref{Fig:8}a and \ref{Fig:8}b at $T>T_c$ one can see  a significant growth of the normal state resonance linewidth. 
 Given the fact the in the experiment\cite{yao2018probe} with FMR in FI/S multilayers the resonance is well-defined at $\Omega \approx 0.01 T_c$ one can conclude that the coupling parameter is $C\sim 0.01$ corresponding to the Fig.\ref{Fig:8}a. In this case there is no noticeable shift of the FMR resonance line 
 as a function of temperature. 
 
 As follows from its definition (\ref{Eq:CouplingParameter}) the coupling parameter $C \propto (d_Fd_M)^{-1}$ can be increased by decreasing either the thickness of the  metal film  $d_M$ or the ferromagnetic film $d_F$. 
 By doing so and reaching the value of $C=0.1$ one would be able to see that the  superconducting correlations produce significant shifht oincrease of the temperature dependence of the resonant field $H_0$.

  \begin{figure}
 \centerline{
 $  \begin{array}{c}
  \includegraphics[width=0.48\linewidth]
 {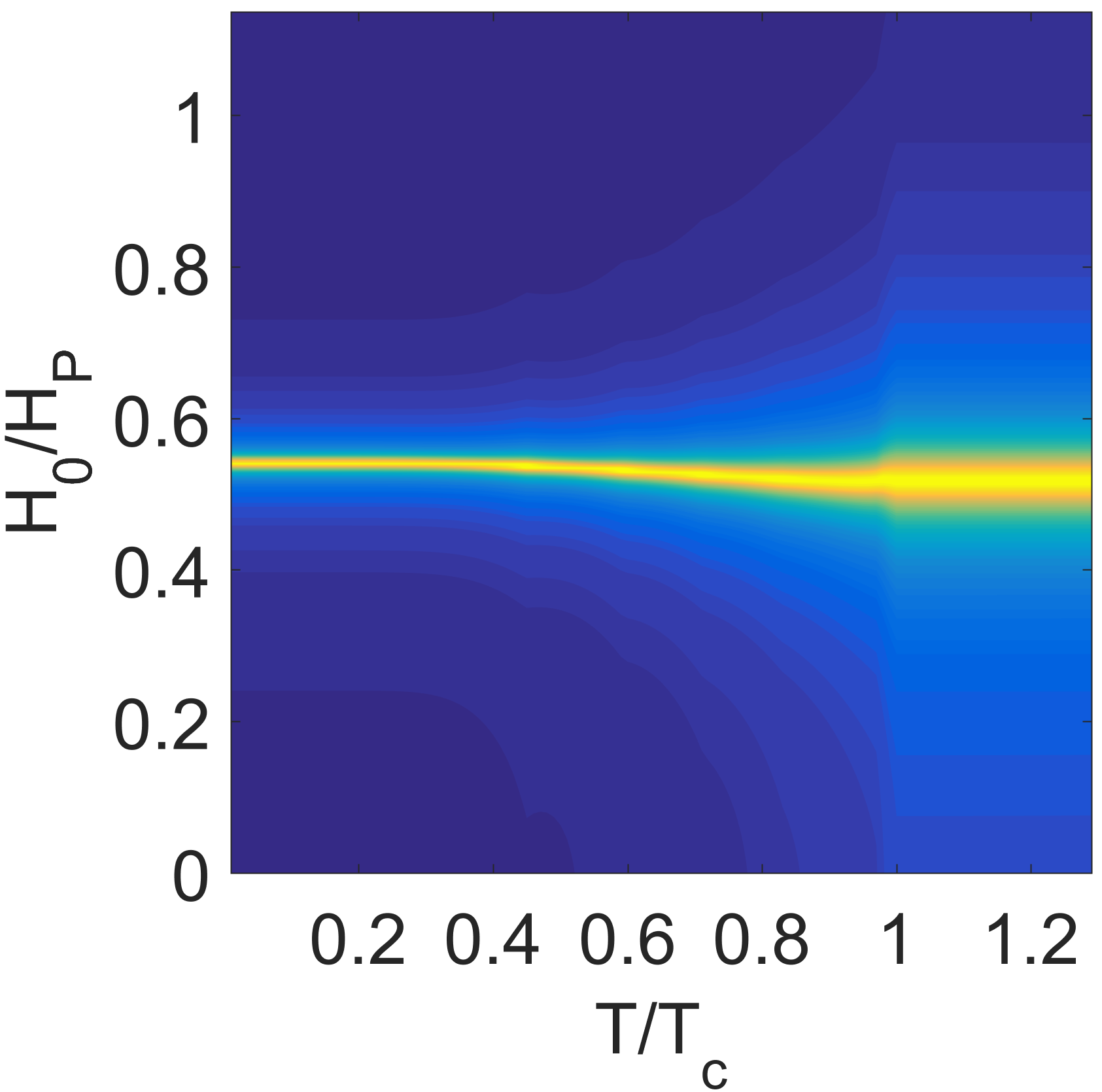} 
 \put (-80,120) { \large{\color{black}  \bf (a) $ \bm {C=0.01}$ }} 
 \;\;  
 \includegraphics[width=0.52\linewidth]
 {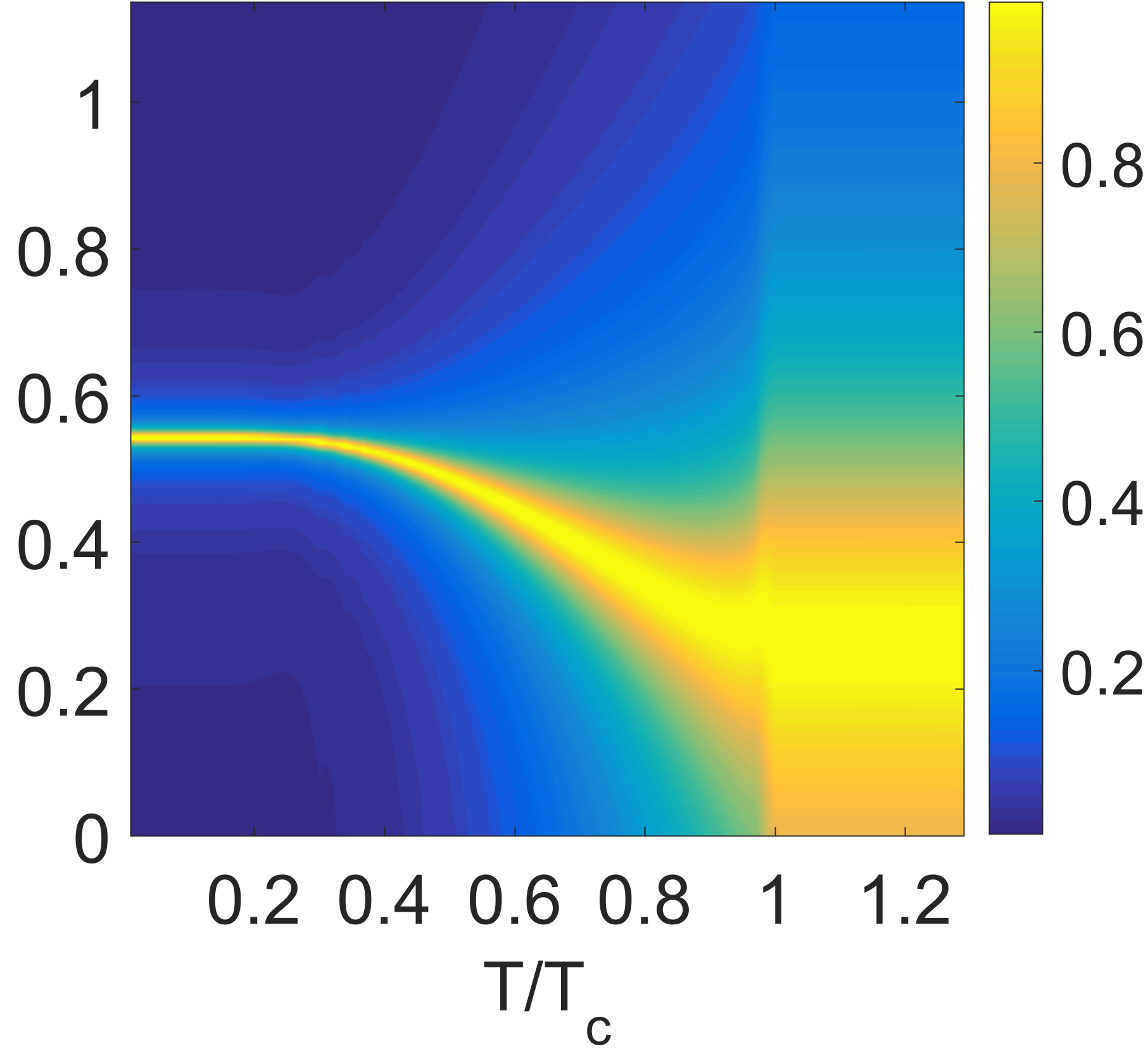} 
 \put (-100,120) {\large{\color{black} \bf (b) $\bm {C=0.1}$ }} 
 \end{array}$}
  \caption{\label{Fig:8} Normalized amplitude of the FMR response signal
    as a function of the constant external magnetic field $H_0$  
   and temperature $T$. The magnetic field is measured in the units $H_p = \Delta (T=0)/\gamma$.  The spin relaxation time is  $\tau_{so}T_c =1 $ and the frequency is $\Omega= T_c$.  We consider (a) weak $C=0.0.1$  and (b) relatively large $C=0.1$ values of the interfacial coupling parameter (\ref{Eq:CouplingParameter}).    
    }
 \end{figure}

  \section{Conclusion}  
     We have derived and analysed the general expression for the time-dependent linear spin response in the superconductor with spin-orbit relaxation. 
 The homogeneous spin susceptibility is found for any amount of the ordinary disorder. In the spatially-inhomogeneous case the diffusive limit is considered.
 We show that the effective spin relaxation rate is given by the sum of  
  the spin-orbit scattering rate and  the diffusive term.
At low frequencies $\Omega\ll T_c$ increasing the effective spin relaxation leads to the formation of the peak in the temperature dependence of the dissipative part of spin susceptibility. This peak is strongly suppressed by increasing the Dynes parameter which models the smearing of the gap edge singularities in the superconductors due to the inhomogeneities or the inelastic phonon scattering.

 Using this result and the model of interfacial exchange interaction 
 we examined the  spin pumping from the 
 ferromagnet with magnetization precession into the adjacent superconducting film. In the low-frequency regime, corresponding to the recent experiments \cite{bell2008spin,jeon2019abrikosov,PhysRevB.99.024507,
 PhysRevApplied.11.014061,Jeon2018,yao2018probe,li2018possible,zhao2020exploring,golovchanskiy2020magnetization} we have analysed the temperature dependence of the additional Gilbert damping parameter induced by the  spin pumping. For realistic values of the Dynes parameter in such materials as NbN this temperature dependence is almost monotonic. This result indicates that there should exist some other 
 mechanism for producing large peaks observed recently in S/FI structures\cite{yao2018probe}. The regime of large Dynes parameters can be also considered to model the spectral smearing which occurs due to the spatial inhomogenuity of the order parameter in systems with metallic ferromagnets. The monotonic suppression of the Gilbert damping parameter in this case corresponds to experimentally observed behaviour of FMR in Py/Nb systems  \cite{bell2008spin}. Similar behaviour is also reproduced by the scattering theory formalism\cite{morten2008proximity}. 
 
 For larger frequencies, comparable with the superconducting gap and enhanced interfacial couplings we get significant shifts 
 of the FMR line. These shifts act towards increasing the resonant field $H_0$ at a given frequency. This behaviour is opposite to the one found in recent  experiments at low frequencies\cite{li2018possible,PhysRevApplied.11.014061, zhao2020exploring,golovchanskiy2020magnetization}. 

\section{Acknowledgements}
This work  was supported by the Academy of Finland (Project No. 297439) and  Russian Science Foundation, Grant No. 19-19-00594.

     \appendix
   \section{Absence of spin response without spin relaxation} 
    \label{Sec:AppNoRel}

 In the absence of spin-orbit scattering $\tau_{so}^{-1} =0 $ and $q=0$ the susceptibility can be written as follows 
  \begin{align} \nonumber
  & \chi_h (\Omega, q=0) = \pi T  \sum_\omega  
  \frac{ \Delta^2 + s_1s_2 - \omega_1\omega_2}
  {s_1s_2 (s_1 + s_2)} 
  \end{align}  
  We can use following relations
  $s_1^2 - s_2^2 = \omega_1^2 - \omega_2^2$
  and $2(\omega_1\omega_2- \Delta^2 -s_1s_2) = 
  (\omega_1+ \omega_2)^2 - 
  (s_1 + s_2)^2 $
  so that
  \begin{align} \nonumber
  &  \sum_\omega 
  \frac{ (\omega_1+ \omega_2)^2 - 
  (s_1 + s_2)^2}
  {s_1s_2 (s_1+s_2)} 
  =
  \\ \nonumber
  &  \sum_\omega \left[
  \frac{ (\omega_1+ \omega_2)^2}
  {s_1s_2 (s_1+s_2)} 
   - 
   \frac{ s_1 + s_2}
  {s_1s_2} \right] =
  \\ \nonumber
  &  \sum_\omega \left[
  \frac{ (\omega_1+ \omega_2) }
  {(\omega_1- \omega_2) } 
  \left( s_2^{-1} - s_1^{-1} \right)
   - 
   s_1^{-1} - s_2^{-1} \right]
   =
    \\ \nonumber
  &- \frac{1}{\Omega}
  \sum_\omega 
   \left[
   (\omega_2- \omega_1) (s_1^{-1} + s_2^{-1}) 
   -
   (\omega_1+ \omega_2) 
   \left( s_2^{-1} - s_1^{-1} \right)
   \right] =
   \\
    &  \frac{2}{\Omega}
  \sum_\omega 
   \left[
     \omega_1 s_1^{-1}  
     -
   \omega_2 s_2^{-1} 
   \right]= \frac{2}{\Omega}
  \sum_\omega 
   \left[
   sgn(\omega_1)
   -
   sgn(\omega_2) \right] 
    \end{align} 
  
  Thus after analytical continuation we can write 
 \begin{align} \nonumber
 \chi_h (\Omega) + 1 =  \int_{-\infty}^\infty \frac{d\varepsilon}{2\Omega}
 [ n_0(\varepsilon+\Omega) - n_0(\varepsilon) ] =   1 
 \end{align}  
  so that $\chi_h (\Omega)=0$ at $\Omega \neq 0$.

 \section{Normal metal limit} \label{AppSec:NormalMetal}
 In the normal metal limit $\Delta=0$ and $\xi_{1,2}= |\omega_{1,2}|$. Then
 \begin{align}
 & \chi_h +1 = \pi T  \sum_\omega  
 \frac{1- sign(\omega_1) sign(\omega_2) }
 {( |\omega_1| + |\omega_2| + 4/3\tau_{so} )}
 \end{align}
 
  Analytical continuation is implemented as follows 
    \begin{align} \nonumber
  & \chi_h +1 = 
  \\ \nonumber
  & \int_{-\infty}^{\infty} \frac{d\varepsilon}{4 i} 
    \frac{ [ n_0 (\varepsilon-1) - n_0 (\varepsilon)] 
    [1-sign(\omega_1)^R sign(\omega_2)^A ]}
  {( |\omega_1|^R + |\omega_2|^A + 4/3\tau_{so} )}  
     \end{align}
  where we have used that
  $|\omega_1|^{R} \to s(-i\varepsilon^{R}_1) = 
   i (\varepsilon - \Omega ) + \Gamma$ and 
  $|\omega_2|^{A} \to s(-i\varepsilon^{A}_2) = 
  - i  \varepsilon + \Gamma $, so that 
  $|\omega_1|^R + |\omega_2|^A \to - i(\varepsilon - \Omega) + i \varepsilon+ 2\Gamma = i\Omega + 2\Gamma$
  
  Then we obtain
  \begin{align} \nonumber
  & \chi_h +1 = 
   \int_{-\infty}^{\infty} \frac{d\varepsilon}{2 i} 
    \frac{[ n_0 (\varepsilon) - n_0 (\varepsilon+\Omega) ]}
  { (   i\Omega + 2\Gamma + 4/3\tau_{so} )} = 
   \\
  & \frac{\Omega}
  {  \Omega - 2i(2/3\tau_{so} + \Gamma)}
  \end{align}
  From this we obtain Eq.\ref{Eq:NormalMetalSusc}.  
  

 \section{Derivation of the strong spin relaxation limit 
  Eq.\ref{Eq:chi_hk0StrongSOExpansion1} }

  Substituting Eq.(\ref{Eq:chi_hk0}) obtained assuming the strong spin relaxation 
  to the general analytical continuation rule 
  (\ref{Eq:AnalyticalContinuationGen}) we obtain 
 \begin{align} \label{Eq:chi_hk0StrongSO}
 & \frac{8}{3 i\tau_{so}} \chi_h =   
    \Delta^2\int  d\varepsilon
   \left [ \frac{F_1(\varepsilon-\Omega)}{\xi^A(\varepsilon)} 
    + \frac{F_1(\varepsilon+\Omega)}{\xi^R(\varepsilon)}
    \right]  
    + 
    \\ \nonumber
   & \int  d\varepsilon
   \left [ \frac{F_2(\varepsilon-\Omega)\varepsilon}
   {\xi^A(\varepsilon)} 
   + 
   \frac{F_2(\varepsilon+\Omega)\varepsilon}
   {\xi^R(\varepsilon)}
   \right] + 
   \\ \nonumber
   & \int  d\varepsilon
    \left [ \frac{F_2(\varepsilon)(\varepsilon+\Omega)}
   {\xi^A(\varepsilon+\Omega)}
   -
     \frac{F_2(\varepsilon-\Omega)\varepsilon}
   {\xi^A(\varepsilon)} 
   \right] 
  \end{align}
 where $\xi^{R,A}(\varepsilon) 
  = \sqrt{ (\varepsilon^{R,A})^2- \Delta^2} $, 
 $F_1 = n_0(\varepsilon)N(\varepsilon)/\varepsilon$, 
 $F_2 = n_0(\varepsilon)N(\varepsilon)$, 
  and $N =  {\rm Re} (\varepsilon /\xi^R) $ is the DOS.  
 The contribution of last term can be calculated 
 to be equal $-i\Omega$ using asymptotic $F_2 (\varepsilon \pm \infty ) = \pm 1$ and $\varepsilon/\xi^A(\varepsilon) \to -1$ at large energies.  
 The first two terms can be calculated using expansions
 $F (\varepsilon \pm \Omega) = F (\varepsilon) \pm \Omega \partial_\varepsilon F$ which yields
 \begin{align} \label{Eq:chi_hk0StrongSOExpansion}
 & \frac{2}{3\tau_{so}}  \frac{{\rm Im} \chi_h}{\Omega} 
 =   
 \int_{-\infty}^{\infty} 
 d\varepsilon
 \frac{N}{\varepsilon}  
 (\Delta^2 \partial_\varepsilon F_1 + 
 \varepsilon \partial_\varepsilon F_2) -1
 \end{align}   
 Integrating by parts this equation can be rewritten as 
 Eq.\ref{Eq:chi_hk0StrongSOExpansion1} in the main text.


 \section{Calculation of  local spin susceptibility in the film of finite thickness} \label{SecApp:CalculationGiniteThickness}

  To take into account finite metallic film thickness
 we incorporate the interfacial exchange field as the 
 boundary conditions to the non-stationary Usadel equations 
  \begin{align} \label{Eq:BCFS}
  D \check g
 \circ \partial_z \check g (z=0) = iJ_{sd}[\bm \sigma \bm m , \hat g]_t 
 \end{align}
   Mathematically it is more convenient to consider the equivalent problem 
 incorporating the interfacial exchange field as the point source to the Usadel equation 
 \begin{align} \label{EqApp:UsadelDeltaz}
 & -i\{\hat\tau_3\partial_t, \check g \}_t  +
 D\partial_z  ( \check g \circ \partial_z \check g) 
 = 
 \\ \nonumber
 & i [\hat\tau_3\hat \tau_2\Delta , \check g ] +
 [\check \Sigma_{so}\circ, \check g]_t 
 + iJ_{sd} \delta (z)[\hat m, \check g ]_t 
 \end{align}
 This equation is considered in the interval $|z|<d_M$. 
 In case if at $z= \pm d_M$ are the interfaces with vacuum 
  the current vanishes   
 \begin{align} \label{Eq:BCvacuum}
 \check g
 \circ \partial_z \check g (z= \pm d_M) = 0
 \end{align}
 In case if at $z= d_M$ are the interfaces with very strong spin sink the 
 correction to GF  vanishes   
 \begin{align} \label{Eq:BCsink}
 \check g_h (z= \pm d_M) =0 
 \end{align}

 We assume that magnetization  depends on time as 
 $\bm m(t) = \bm m_\Omega e^{-i\Omega t}$ and search for the corrections to the  GF in the form 
 \begin{align}
 & \hat g (t,t^\prime) =  T \sum_\omega 
 [\hat g_0 (1) e^{-i\omega_1 (t_1-t_2)} + 
 \hat g_h (12) e^{-i(\omega_1 t_1-\omega_2 t_2)} ]
 \end{align}
where $\omega_2 = \omega_1 - \Omega$
and $\hat g_h$  represents the correction to the first order of the oscillating field $\bm m_\Omega$. 
To satisfy boundary conditions we search the solution in the form 
\begin{align}
    \hat g_h (12) = \sum_{n=0}^{\infty} g_{q_n}(12) \cos(q_n z)
\end{align}
with $ q_n=n \pi /d_M$ in case of the vacuum interface (\ref{Eq:BCvacuum}) 
and $q_n = (n+1/2)\pi/d_M$ in case of the strong spin sink interface (\ref{Eq:BCsink}) . 
Using the expansion 
$\delta (z) =(2d_M)^{-1} \sum_{n} \cos(q_n z) $
We have the equation for the correction
 \begin{align} \label{SMEq:}
 & (\tilde s_1  + D q^2 ) \hat g_0 (1) \hat g_q(12) - \tilde s_2 \hat g_q(12) \hat g_0 (2) =  
 \\ \nonumber
 & i (\bm h_\Omega\bm{\hat \sigma}) 
 [\hat g_0 (1) \hat \tau_3
 - 
 \hat \tau_3 \hat g_0 (2)] 
 \end{align}
 where $\bm h_\Omega = (G_{i}^{\uparrow\downarrow}/2\nu d_M) \bm m_\Omega$. Using the commutation relation 
 $ \hat g_0 (1) \hat g_k (12) +  \hat g_k (12)\hat g_0 (2)=0$ we get the solution is given by 
 \begin{align}\label{SMEq:gk12Solutions}
 \hat g_q (12) = i (\bm h_\Omega\bm{\hat \sigma})
 \frac{\hat\tau_3 - 
 \hat g_0(1) \hat\tau_3 \hat g_0 (2)}
 {s_1+s_2 + 4/3\tau_{so} + D q^2 }
 \end{align}
     
The spin polarization at the M/F interface which can be written in terms if the susceptibility 
  \begin{align}
  & \bm S(z=0) = \nu h_{eff}\chi_m(\Omega) \bm m_\Omega 
   \end{align}
  Substituting the solution (\ref{SMEq:gk12Solutions}) to the expression for the spin polarization 
 \begin{align} \label{SMEq:SpinDensityOnShell}
 \bm S (t,z) =  -i\frac{\pi \nu}{4} {\rm Tr} [ 
 \bm{\hat \sigma} \hat\tau_3 \hat g]|_{ t_{1,2}=t}.
 \end{align}
 we get the imaginary frequency local susceptibility of the finite-thickness film (\ref{Eq:ChiM_def}) .

  \bibliography{refs2}

 \end{document}